\documentclass[aps,noshowpacs,superscriptaddress,twocolumn,floatf ix]{revtex4}

\pdfoutput=1
\usepackage[utf8]{inputenc}
\usepackage[english]{babel}
\usepackage[T1]{fontenc}
\usepackage{braket}
\usepackage{natbib}
\usepackage{tikz}
\usepackage{fixltx2e} 
\usepackage{dcolumn}
\usepackage{pgfplots}
\usepackage{lipsum}
\usepackage{booktabs}
\usepackage{subcaption}
\usepackage{pgfplots}
\usepackage{amsmath,amssymb,graphicx}

\usepackage{adjustbox}
\usepackage{hyperref}

\makeatletter

\newtheorem{theorem}{Identity}[section]

\newcommand{\norm}[1]{\left\lVert#1\right\rVert}
\newcommand{\R}{\mathbb{R}}

\def\ket#1{\mathinner{|{#1}\rangle}}

\usepackage[normalem]{ulem}

\pgfplotsset{compat=1.17}

\begin{document}

\title{Quantum Fourier Networks for Solving Parametric PDEs}

\author{Nishant Jain}
\affiliation{QC Ware, Palo Alto, USA and Paris, France}
\affiliation{Indian Institute of Technology, Roorkee, India}

\author{Jonas Landman}
\affiliation{QC Ware, Palo Alto, USA and Paris, France}
\affiliation{School of Informatics, University of Edinburgh, Scotland, UK}

\author{Natansh Mathur}
\affiliation{QC Ware, Palo Alto, USA and Paris, France}
\affiliation{IRIF, CNRS - Université Paris-Cité, France}

\author{Iordanis Kerenidis}
\affiliation{QC Ware, Palo Alto, USA and Paris, France}
\affiliation{IRIF, CNRS - Université Paris-Cité, France}


\begin{abstract}
Many real-world problems, like modelling environment dynamics, physical processes, time series etc., involve solving Partial Differential Equations (PDEs) parameterised by problem-specific conditions. Recently, a deep learning architecture called Fourier Neural Operator (FNO) proved to be capable of learning solutions of given PDE families for any initial conditions as input. However, it results in a time complexity linear in the number of evaluations of the PDEs while testing. Given the advancements in quantum hardware and the recent results in quantum machine learning methods, we exploit the running efficiency offered by these and propose quantum algorithms inspired by the classical FNO, which result in time complexity logarithmic in the number of evaluations and are, therefore, expected to be substantially faster than their classical counterpart. At their core, we use the unary encoding paradigm and orthogonal quantum layers and introduce a circuit to perform quantum Fourier transform in the unary basis. We propose three different quantum circuits to perform a quantum FNO. The proposals differ in their depth and their similarity to the classical FNO.
We also benchmark our proposed algorithms on three PDE families, namely Burgers' equation, Darcy’s flow equation and the Navier-Stokes equation. The results show that our quantum methods are comparable in performance to the classical FNO. We also perform an analysis on small-scale image classification tasks where our proposed algorithms are at par with the performance of classical CNNs, proving their applicability to other domains as well.
\end{abstract}

\maketitle

\section{Introduction}

\subsection{Fourier Neural Network}

\begin{figure*}[]
    \centering
    \includegraphics[width=\linewidth]{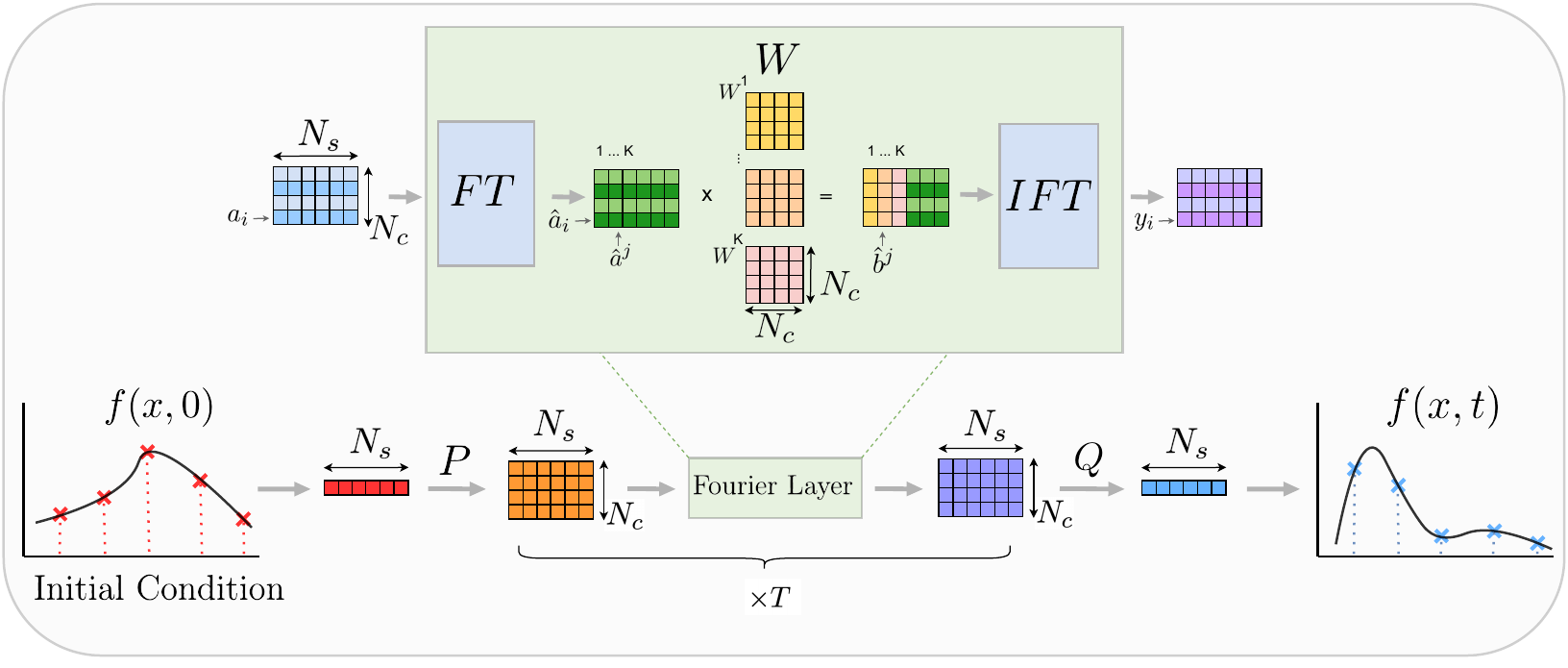}
    \caption{\textbf{Overview of the Fourier Neural Network.} Each initial condition $f(x,0)$ is sampled $N_s$ times and modified via a trainable matrix $P$ to become a matrix of size $N_s \times N_c$. Then, $T$ Fourier Layers (green block) are applied sequentially. In this paper, we are designing quantum circuits to implement the Fourier Layers.
    Each Fourier Layer consists of a row-wise Fourier Transform (FT), followed by a column-wise multiplication with trainable matrices labelled $W$, and the row-wise Inverse Fourier Transform (IFT). We only apply the inner matrix multiplications to the first $K$ columns (also called \emph{modes}) and leave the others unchanged or replaced by 0s before the IFT.
    Finally, a reverse operation with a trainable matrix $Q$ is performed to obtain the output $f(x,t)$ as a discretised vector. The trainable parts are updated with Gradient Descent until the outputs correspond to the actual solutions of the PDE.}
    \label{fig:classical_FNO_fig1}
\end{figure*}

Solving Partial Differential Equations (PDEs) has been a crucial step in understanding the dynamics of nature. They have been widely used to understand natural phenomena such as heat transfer, modelling the flow of fluids, electromagnetism, etc. 

Each PDE is an equation, along with some initial conditions, for which the solution is a function $f$ of space and time $(x,t)$, for instance. 
A \emph{PDE family} is determined by the equation itself, such as Burgers' equation or Navier-Stokes equation. An \emph{instance} of a given PDE family is the aforementioned equation along with a specific initial condition, represented, for instance, as $f(x,t_0)$. Modifying this initial condition leads to a new PDE instance and, therefore, to a new solution $f(x,t)$. Note also that the solution is highly dependent on some physical parameters (\emph{i.e.} viscosity in fluid dynamics).

In practical scenarios, a closed-form solution for most PDE families' instances is difficult to find. Therefore, classical solvers often rely on discretising the input space and performing many approximations to model the solution. A large number of computations for each PDE instance are required, depending on the chosen \emph{resolution} of the input space.

Recently, considerable effort in research for approximating a PDE's solution is based on neural networks. 
The main idea is to let a neural network become the solution of the PDE by training it either for a fixed PDE instance or with various instances of a PDE family. The network is trained in a supervised way by trying to match the same solutions as the ones computed with classical solvers.
The first attempts \cite{yu2018deep, bar2019unsupervised} were aimed at finding the PDE's solution $f(x,t)$ for an input $(x,t)$ given a specific initial condition (one PDE instance), and later \cite{zhu2018bayesian, adler2017solving, bhatnagar2019prediction} to a specific discretisation resolution for all instances of a PDE family. For the first case, once trained, the neural network can output solution function values at any resolution for the instance it was trained on. However, it has to be optimised for each instance (new initial condition) separately. In the latter case, the neural network can predict solution function values for any instance of the PDE family but for a fixed resolution on which it was trained.



A recent proposal named \emph{Fourier Neural Network} \cite{li2020fourier} overcame these limitations and posed the problem as learning a function-to-function mapping for \emph{parametric} PDEs. Parametric PDEs are families of PDEs for which the initial condition can be seen as parametric functions. 
Given any initial condition function of one such PDE family sampled at any resolution, the neural network can predict the solution function values at the sampled locations.

The input is usually the initial condition $f(x,t_0)$ itself. It is encoded as a vector of a certain length $N_s$ by sampling it uniformly at $N_s$ locations $x$ of the input space, given some resolution. This input is also called an \emph{evaluation} of the initial condition function. The number of samples $N_s$ is key in analysing the computational complexity as it is the neural network input size. Note that sometimes the initial condition is also sampled for several times $t$ as well.
The output of the neural network is the corresponding PDE's solution $f(x,t)$ applied at all $x$ sampled and for a fixed $t$.
Experiments on widely popular PDEs showed that it was effective in learning the mapping from a parametric initial condition function to the solution operator for a family of PDEs.

The method proposes a \emph{Fourier Layer} repeated several times. It consists of a Fourier Transform (FT), then a multiplication with a trainable matrix (also called a \emph{linear transform}, and an Inverse Fourier Transform (IFT) operation, and ends with a standard non-linearity. This is similar to the convolution operation as it also translates to multiplication in the Fourier domain. However, a key feature of the Fourier Layer is the fact that one can keep only some part of the data before the IFT, corresponding to the lowest frequency in the Fourier domain, reducing the amount of information and computational resources.

The major bottleneck which might hinder the scalability of this classical Fourier Neural Operator (FNO) is its time complexity, limited by the classical FT and IFT operations inside the Fourier Layer. Indeed, their time complexity is $O(N_s \log N_s)$ with a classical computer, where $N_s$ is the input size (number of samples). In many use cases, $N_s$ is expected to be quite high for learning precisely the solution of a PDE family. To be more precise, we will see that each input, a vector of size $N_s$, is first modified and reshaped to become a matrix of size $N_s \times N_c$ (see Fig.\ref{fig:classical_FNO_fig1}). $N_c$ is named \emph{channel dimension}, and usually $N_s \gg N_c$. This matrix will be the actual input of the Fourier Layer.
\subsection{Quantum Algorithmic Proposals}

Quantum computing has gained popularity due to its potential for faster performance than its classical counterpart. Among the most famous quantum algorithms is the exponentially faster Quantum Fourier Transform (QFT), although probably only attainable with long-term quantum computers. In the same way, long-term quantum machine learning and deep learning were proposed \cite{allcock2020quantum, kerenidis2019quantum, berner2021quantum}.

More recently, several developments in learning techniques based on near-term quantum computing were proposed.
The initial demonstrations of these algorithms involved experiments on small-scale quantum hardware \cite{farhi2018classification, coyle2020born, cappelletti2020polyadic, grant2018hierarchical}, which established their effectiveness in extracting patterns. Following this, many works \cite{mari2020transfer, beer2020training} proposed small-scale implementations of fully connected quantum neural networks on near-term hardware. Other proposals \cite{cong2019quantum} for deploying convolution-based learning methods on quantum devices showed effective training in practice. Furthermore, \cite{chakrabarti2019quantum} proposed quantum-hardware implementation for generative adversarial networks. A different approach, where the inputs are encoded as unary states, using the two-qubit quantum gate RBS (\emph{Reconfigurable Beam Splitter}) was proposed in a recent work \cite{johri2021nearest}. This encoding gave rise to the use of orthogonal properties of pure quantum unitaries, as proposed in \cite{landman2022quantum, cherrat2022quantum} for training, for instance, orthogonal feed-forward networks to damp the gradient-based issues while learning. It used a pyramid-shaped (or other architectures) circuit based on parameterised RBS gates to implement a learnable orthogonal matrix as compared to the existing classical approaches, which offer approximate orthogonality at the cost of increased training time. This orthogonality in neural networks results in much smoother convergence and also fewer parameters, as shown by \cite{li2019orthogonal} for feed-forward neural networks and \cite{wang2020orthogonal} for convolutional nets.
The effectiveness of these orthogonal quantum networks was further shown in another work on medical image classification \cite{landman2022quantum} problem. 


In this work, we develop quantum algorithms to implement the Fourier Neural Operator (FNO). In particular, we propose three circuits equivalent to or close to the Fourier Layer (See the green box in Fig.\ref{fig:classical_FNO_fig1}). The remaining parts of the FNO can be adapted using existing techniques \cite{landman2022quantum}.

At the core of our circuit, we develop a new Quantum Fourier Transform (QFT) suited for near-term hardware and specific quantum data encoding. Termed as \textit{unary-QFT}, which transforms only the \emph{unary} states into the Fourier domain, with an exponential speedup compared to the classical operation. 
We have built it upon the recently developed idea \cite{landman2022quantum,johri2021nearest} of encoding inputs to unary quantum states and applying orthogonal transformations only on these unary-basis states via learnable quantum circuits. 
Using this, we adapt the classical Fourier Neural Network \cite{li2020fourier} and propose several quantum algorithms to learn the functional mapping from the initial condition function of a PDE instance to the corresponding solution function.

The circuit proposed in this work is inspired by the widely popular butterfly diagram used for the Fast Fourier Transform (FFT) \cite{cooley1965algorithm}. Then, 
we propose an implementation of controlled butterfly-shaped learnable quantum circuits for applying the linear transform (trainable matrix multiplications) in the Fourier domain.
This results in three quantum circuits inspired by the classical Fourier Layer, which are faster than the classical operation or, said differently, require fewer parameters for the same architecture, thereby boosting their scalability. 
Given the matrix input of the Fourier Layer dimension $N_s\times N_c$, where $N_s$ corresponds to the number of samples per PDE, and $N_c$ correspond to the channel dimension, the order of time complexity corresponding to Fourier Layer and proposed algorithms is shown in table \ref{tab:complexity}.

\begin{table*}[!htb]
   \begin{center}
    \resizebox{\linewidth}{!}{
    \begin{tabular}{lcccc}
    \toprule
    
    Method   & \#Qubits & \#Circuits & Gate Complexity & Depth Complexity \\
    \midrule
        Classical FL  & - & - & -& $N_c$+$N_s$\text{log}($N_s$)\\
        Parallel QFL   & $N_c$ + $N_s$  & $K$ & $KN_c$\text{log}($N_c$)+$KN_cN_s$\text{log}($N_s$) & $N_c$+$N_c$\text{log}($N_s$)\\
        \midrule
        Sequential QFL   & $N_c$ + $N_s$  & 1 & $KN_c$\text{log}($N_c$)+$N_cN_s$\text{log}($N_s$)& $KN_c$+$N_c$\text{log}($N_s$) \\
        Composite QFL   & $N_c$ + $N_s$ & 1 & ($N_c$+$K$)\text{log}($N_c$+$K$)+$N_cN_s$\text{log}($N_s$) & \text{log}($N_c$+$K$)+$N_c$\text{log}($N_s$)  \\
    \bottomrule
    \end{tabular}
    }
\end{center}
\vskip -0.15in
\caption{ Comparison of the order of time/depth complexities (Big-$O$) of the proposed circuits with the existing classical Fourier Layer (FL). Here $N_s$ denotes the sampling dimension, $N_c$ denotes the channel dimension where $N_s\gg N_c$ and $K$ (usually in the range 4-16) denotes the maximum number of modes allowed \cite{li2020fourier}. This implies that the proposed quantum algorithms would be faster than the classical method. Each quantum circuit requires $N_c+N_s$ qubits and $K$ independent parallel circuits are required by the Parallelised QFNO.
    }
    \label{tab:complexity}
\end{table*}

The first algorithm corresponds to the quantum counterpart of the classical operation, where the middle trainable matrix is an orthogonal one. The other two algorithms are modifications of the first circuit with lower circuit depth (at the cost of more qubits or different expressivity) to mitigate the fact the near-term quantum hardware might still be too noisy. 

We have simulated all the three proposed quantum algorithms on all the three PDEs evaluated, as in the classical FNO paper \cite{li2020fourier}, namely Burgers' equation, Darcy's flow equation and Navier-Stokes equation, on the synthetic datasets used in that paper. We have also simulated our quantum algorithms against the Convolutional Neural Networks (CNNs) on several benchmark datasets for image classification. 
In all the experiments, the three quantum algorithms perform similarly and also show accuracies comparable to state-of-the-art classical methods. 

\subsection{Contributions}

To summarise, the contributions of this work are the following:
\begin{itemize}
    \item We propose a new quantum circuit for performing a Quantum Fourier Transform on an input encoded as a superposition of unary states.
    \item Using the above-mentioned unary-QFT, we propose three quantum circuits with a provable equivalence or approximation to the classical Fourier Layer. These circuits can be sequentially combined to form a quantum version of a trainable Fourier neural network for solving Parametric PDEs.
    \item We provide an in-depth analysis of the computational complexity of the circuits and prove their logarithmic time complexity with respect to $N_s$ (input dimension), compared to the linear time complexity of their classical counterpart.
    \item We benchmark results, showing the effectiveness of the proposed quantum algorithms, against their classical equivalent, for solving several PDEs or even classifying images from well-known datasets.
\end{itemize}

\section{Classical Fourier Neural Operator}
\label{sec:classical_fno}
For solving a Partial Differential Equation (PDE), we are provided with a dataset where each instance is a set of the initial conditions of the family of the PDE. This initial condition is represented as a parameterised function and is sampled at various locations to generate the input. The neural network's output corresponding to this initial condition is trained to be the value of the solution $f(x,t)$ for that instance at the same locations and for a given $t$. The Fourier Neural Operator (FNO) \cite{li2020fourier} tries to learn this functional mapping from the initial condition function to the solution function for this PDE family. This implies that given a set of initial conditions sampled at various locations as input to the FNO, it has to predict the solution function values at all these locations for any PDE instance in the test set. An overview of the FNO is given as Fig.\ref{fig:classical_FNO_fig1}.

We recall that each input of the FNO is a parametric initial condition, seen for instance as a function of space $f(x,0)$. This function is sampled at $N_s$ locations and forms a vector in $\mathbb{R}^{N_s}$ that is later modified by a trainable matrix $P$ to become a matrix $A \in \mathbb{R}^{N_c \times N_s}$. We denote $N_c$ the \emph{channel dimension} and usually $N_s \gg N_c$. The matrix $A$ will be the input of the first \emph{Fourier Layer} (FL) and its size will be maintained for each following FL. As shown in Fig.\ref{fig:classical_FNO_fig1}, the FNO consists of a sequence of Fourier Layers. Without loss of generality, we will denote $A \in \mathbb{R}^{N_c \times N_s}$ to be the input of any Fourier Layer.


Each Fourier Layer (FL) starts by transforming its input matrix to the Fourier domain. It applies a Fourier Transform (FT) on each column of the input matrix $A$. The resulting matrix has the same dimension, and we will refer to its $N_s$ columns as \emph{modes} to emphasize their presence in the Fourier basis.
After the FT, we apply a learnable matrix multiplication to the first $K$ modes, \emph{i.e.} the first $K$ columns among $N_s$ (See Fig.\ref{fig:classical_FNO_fig1}). 
The original proposal indicates to \emph{crop} the remaining modes by replacing them with 0. In our quantum proposal, we will let the remaining modes untouched rather than cropping them. The final operation is an Inverse Fourier Layer (IFT) which transforms the matrix back to the input space. Note that in the original proposal, the authors apply in parallel a direct convolution to the input, and both outputs are merged (Not shown in Fig.\ref{fig:classical_FNO_fig1}. In our quantum proposals, we discard this convolution part for simplicity, without any impact on the experimental accuracy. 

In this work, we propose several quantum algorithms for mimicking the Fourier Layer and name these circuits \textit{Quantum Fourier Layer} (QFL). We name the resulting neural network as a \textit{Quantum Fourier Neural Operator} (QFNO). As the other parts of the FNO ($P$ and $Q$ matrix multiplications from Fig.\ref{fig:classical_FNO_fig1}) are easier to adapt with existing quantum techniques \cite{landman2022quantum}, we have not focused our work on them. Next, we will formulate the analytic expression of the FL's output, in order to prove its correspondence to QFL. 




\noindent
\textbf{Fourier Layer}    
We now discuss the mathematical details of the classical Fourier Layer for a 1D PDE case (\emph{e.g.} Burgers' equation),  showing the inputs and outputs of each transformation involved. For each PDE instance, the input is denoted as $A \in \R^{N_c\times N_s}$, where $N_c$ denotes the number of channels per sample in the input and $N_s$ corresponds to the number of samples for this instance (initial condition function). 

Regarding notation, the elements of $A$ are $a_{ij}$, its rows are $a_i$ while its columns are $a^j$. We denote the output corresponding to this classical operation as $Y \in \mathbb{R}^{N_c\times N_s}$, and its elements, columns, and rows as $y_{ij}$, $y_i$, and $y^j$ respectively. 
As the quantum matrices are orthogonal and the $l_2$-norm of any quantum state vector is 1, we consider the input $A$ such that $||A||_2=1$. Enforcing this condition is easy and doesn't have any significant impact on the optimisation process.
Going further, a Fourier Transform (FT) is applied to this input along each row of size $N_s$: 
\begin{equation}
    \hat{a}_{i} := FT(a_i)
\end{equation}    
where $a_i = (a_{ij})_{j\in [1,N_s]}$. We can also define $\hat{a}^{j}=(\hat{a}_{ij})_{i\in[1,N_c]}$. Let $\hat{A}$ be the resulting matrix.

Denoting the maximum number of modes with $K$, the intermediate linear transform is in fact a multiplication with a 3-tensor $W \in \mathbb{R}^{N_c\times N_c \times K}$. Each $W^{k}\in \mathbb{R}^{N_c\times N_c}$ corresponds to the $k^{th}$ matrix of $W$, indexed along the last dimension, and corresponding to $k^{th}$ mode (see Fig.\ref{fig:classical_FNO_fig1}). In the quantum implementation later, we will consider matrices $W^k$ to be  orthogonal, as this naturally occurs in quantum circuits. We multiply the tensor $W$ to the first $K$ modes (along the $N_s$ dimension) of the $\hat{A}$. Said differently, for each $j\leq K$, the $j^{th}$ column of $\hat{A}$ is multiplied by $W^j$, resulting in the following output

\begin{equation}
    \left[(W^{j}\hat{a}^{j})_{j\in[1,K]} ,  (\hat{a}^j)_{j\in[K+1,N_s]}\right]
\end{equation}

Let $\hat{b}^j = W^j \hat{a}^j$, we can rewrite the previous vector as 

\begin{equation}
    \left[(\hat{b}^{j})_{j\in[1,K]} ,  (\hat{a}^{j})_{j\in[K+1,N_s]}\right]
    \label{eq:classical_before_iqft}
\end{equation}

In the original classical proposal \cite{li2020fourier}, the rest of the modes are discarded (replaced by zeros). In the quantum case, it will be simpler to let the other modes unchanged. We found that this choice doesn't impact the performance.


Finally, we apply the Inverse Fourier Transform ($IFT$) operation on this transformed input, row by row. It results in the following output for each row $i$ :
\begin{equation}
    y_i = 
    IFT\left(\left[(\hat{b}_{ij})_{j\in[1,K]} ,  (\hat{a}_{ij})_{j\in[K+1,N_s]}\right]\right)
    \label{eq:classical_out}
\end{equation}
Where $\hat{b}_{ij}$ is the $i^{th}$ component of $\hat{b}^j$. In conclusion, the overall time complexity of the Fourier Layer ($FT$ + Matrix Multiplications + $IFT$) should be $O(KN^2_c+2N_cN_slog(N_s))$. 
This runtime can be improved if we consider a distributed algorithm, considering the current availability of efficient GPUs. By parallelising classical operations, we can achieve the Fourier Layer in $O(N_c+2N_slog(N_s))$. Note however that the dominant term remains the same as $N_s \gg N_c$ and $K$ is usually a constant.


\section{Quantum Algorithmic Tools}\label{sec:quantumtools}

In this section, we introduce quantum tools necessary to build the Quantum Fourier Layer in Section \ref{sec:quantumFourierLayer}. These tools are meant to be implemented on near-term quantum computers, with modularity so that they can be useful for other applications.

We first introduce the matrix unary amplitude encoding, a fast way to load a matrix as a quantum state. 
Then, we develop a new quantum circuit to apply a Quantum Fourier Transform (QFT) on the unary basis states. 
Finally, we present learnable quantum orthogonal layers, the equivalent of learnable weight matrices in classical neural networks.

We describe here a quantum gate that will be common to the next tools: the 2-qubit Reconfigurable Beam Splitter $(RBS)$ gate \cite{foxen2020demonstrating},
parameterised by a single parameter $\theta$. The $RBS$ gate has the following unitary:
\begin{equation}
RBS (\theta) = 
\begin{pmatrix}
1 & 0 & 0 & 0\\
0 & \cos(\theta) & \sin(\theta) & 0\\
0 & -\sin(\theta) & \cos(\theta) & 0\\
0 & 0 & 0 & 1\\
\end{pmatrix}
\end{equation}\label{eq:RBS}

It can be observed that it modifies $\ket{01}$ and $\ket{10}$, while it performs the identity operation on $\ket{00}$ and $\ket{11}$. The $RBS$ gate, therefore, preserves the hamming weight of the input state. In particular, any superposition of the unary basis (states with hamming weight 1), is kept in this basis through a circuit made of $RBS$ gates. Its implementation depends on the quantum hardware considered.




Now, we will discuss an identity for these RBS gates  and use it in the coming section to implement controllable parameterised circuits made of these gates.

\begin{figure}[!htb]
    \centering
    \includegraphics[width=\linewidth]{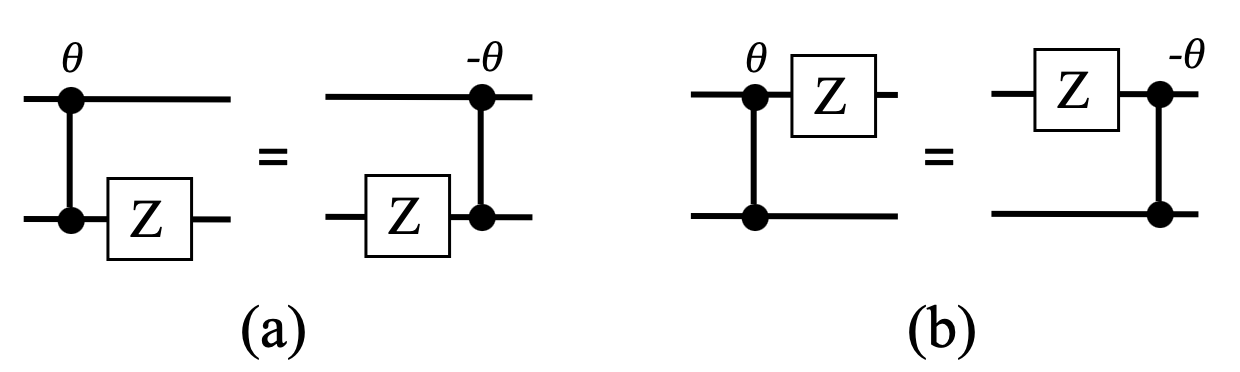}
    \caption{RBS identities}
    \label{fig:rbs_identity}
\end{figure}

\begin{theorem}
\label{identity_rbs}
   Given two qubits, applying an RBS gate on them with angle $\theta$, followed by a $Z$ gate on any one of them is equivalent to applying a $Z$ gate on the same qubit followed by an RBS gate with angle $-\theta$ on the two qubits (Figure \ref{fig:rbs_identity}). 
\end{theorem}
\textbf{Proof}: Let us first look at the circuit shown in Figure \ref{fig:rbs_identity} (a). Eq. \ref{eq:rbs_id_1} shows the calculation for the final unitary corresponding to the left-hand side: 
\\
\begin{equation}
\begin{aligned}
       & \begin{pmatrix}
        1&0&0&0 \\
        0&-1&0&0 \\
        0&0&1&0 \\
        0&0&0&-1 \\
    \end{pmatrix} 
    \cdot
    \begin{pmatrix}
        1&0&0&0 \\
        0&cos(\theta)&sin(\theta)&0 \\
        0&-sin(\theta)&cos(\theta) &0 \\
        0&0&0&1 \\      
    \end{pmatrix} \\
&=
    \begin{pmatrix}
        1&0&0&0 \\
        0&-cos(\theta)&-sin(\theta)&0 \\
        0&-sin(\theta)&cos(\theta) &0 \\
        0&0&0&-1 \\      
    \end{pmatrix} 
    \end{aligned}
 \label{eq:rbs_id_1}
\end{equation}

\noindent
and Eq. \ref{eq:rbs_id_2} shows the same for the right-hand side:

\begin{equation}
\begin{aligned}
        &\begin{pmatrix}
        1&0&0&0 \\
        0&cos(\theta)&-sin(\theta)&0 \\
        0&sin(\theta)&cos(\theta) &0 \\
        0&0&0&1 \\      
    \end{pmatrix} 
    \cdot
    \begin{pmatrix}
        1&0&0&0 \\
        0&-1&0&0 \\
        0&0&1&0 \\
        0&0&0&-1 \\
    \end{pmatrix} \\
    &=
    \begin{pmatrix}
        1&0&0&0 \\
        0&-cos(\theta)&-sin(\theta)&0 \\
        0&-sin(\theta)&cos(\theta) &0 \\
        0&0&0&-1 \\      
    \end{pmatrix} 
    \end{aligned}
    \label{eq:rbs_id_2}
\end{equation}

\noindent
where both the equations finally arrive at the same unitary matrix, thereby proving the identity.
A similar calculation for the circuit shown in Figure \ref{fig:rbs_identity} (b) verifies its correctness.

\subsection{Data Encoding in the Unary Basis}\label{sec:dataloader}

As seen in Section \ref{sec:classical_fno}, the input of each classical Fourier Layer is a matrix $A$. As we are about to propose quantum circuits to process this data, we need a method to encode these matrices as quantum states. We chose to encode data as amplitude-encoded states, a superposition of basis states with amplitudes that correspond to the data itself. We chose the unary basis, namely the computational basis vectors that have a hamming weight of 1, e.g. $\ket{e_i} = \ket{0\cdots010\cdots0}$ with the $1$ on the $i^{th}$ qubit. This choice of basis is motivated by its ability to implement near-term encoding for vectors and matrices. It also allows performing tractable linear algebra tasks with provable guarantees. Higher order basis states can be used \cite{kerenidis2022quantum} but are not the focus of this work.

Given an input matrix $A \in \R^{n\times d}$, its quantum state once loaded should be:

\begin{equation}
\label{eq:quantum_matrix_data_state}
    \ket{A} = \frac{1}{\norm{A}}\sum_{i=1}^n\sum_{j=1}^d a_{ij}\ket{e_j}\ket{e_i}
\end{equation}

\begin{figure}
    \centering
    \includegraphics[width=\linewidth]{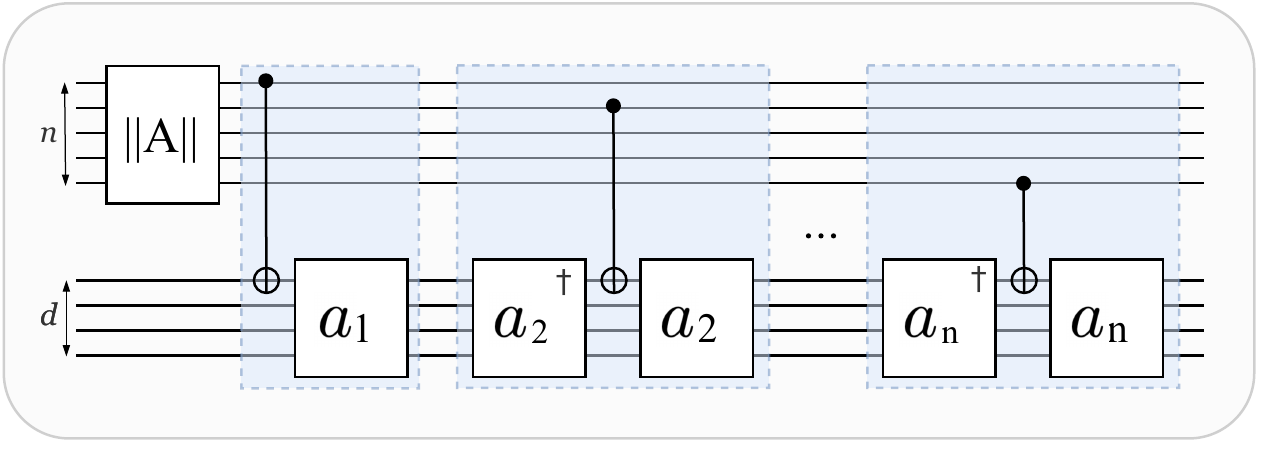}
    \caption{Quantum circuit for loading a matrix $A\in\R^{n\times d}$ as a quantum state on the unary basis. Each white square is a circuit to load a vector. $a_i$ is the $i^{th}$ row of $A$. The circuit starts with the state $\ket{0}^n\otimes\ket{0}^d$}.
    \label{fig:matrixdataloader}
\end{figure}

To load a matrix in such a way, we use a quantum circuit made of two registers (one for the rows, the other for the columns) as shown in Fig.\ref{fig:matrixdataloader} from \cite{cherrat2022quantum}. It uses circuits from \cite{johri2021nearest} that load vectors in the unary basis. For instance, a row $a_i \in \R^d$ is loaded as $\ket{a_i} = \frac{1}{\norm{a_i}}\sum_{i=1}^n a_{ij} \ket{e_i}$. We can get rid of the normalisation factors if we assume or preprocess the vectors and matrices to be normalised.

On the $n$-qubits top register, we first load the vector made of the norm of each row $(\norm{A_1},\cdots,\norm{A_n})$. On the $d$-qubits lower register, we sequentially load and unload each row $A_i$ in a controlled fashion. Details can be found in \cite{cherrat2022quantum}.

With the right connectivity, the circuit to load a vector of size $n$ has depth $O(log(n))$, hence loading a matrix $A\in\R^{n\times d}$ requires a circuit of depth $O(log(d)+2dlog(n))$.

\subsection{Unary Quantum Fourier Transform}\label{sec:unaryQFT}

Quantum Fourier Transform (QFT), one of the most impactful algorithms found in quantum computing literature, provides an exponential speedup compared to classical computing. It performs the Discrete Fourier Transform over the entire $2^n$-dimensional Hilbert space. In this work, we propose a new quantum circuit that performs the Discrete Fourier transform over the unary basis states. This allows for a shallow-depth quantum circuit adapted to our quantum data encoding presented in the previous section.  

\begin{figure}
    \centering
    \includegraphics[width=\linewidth]{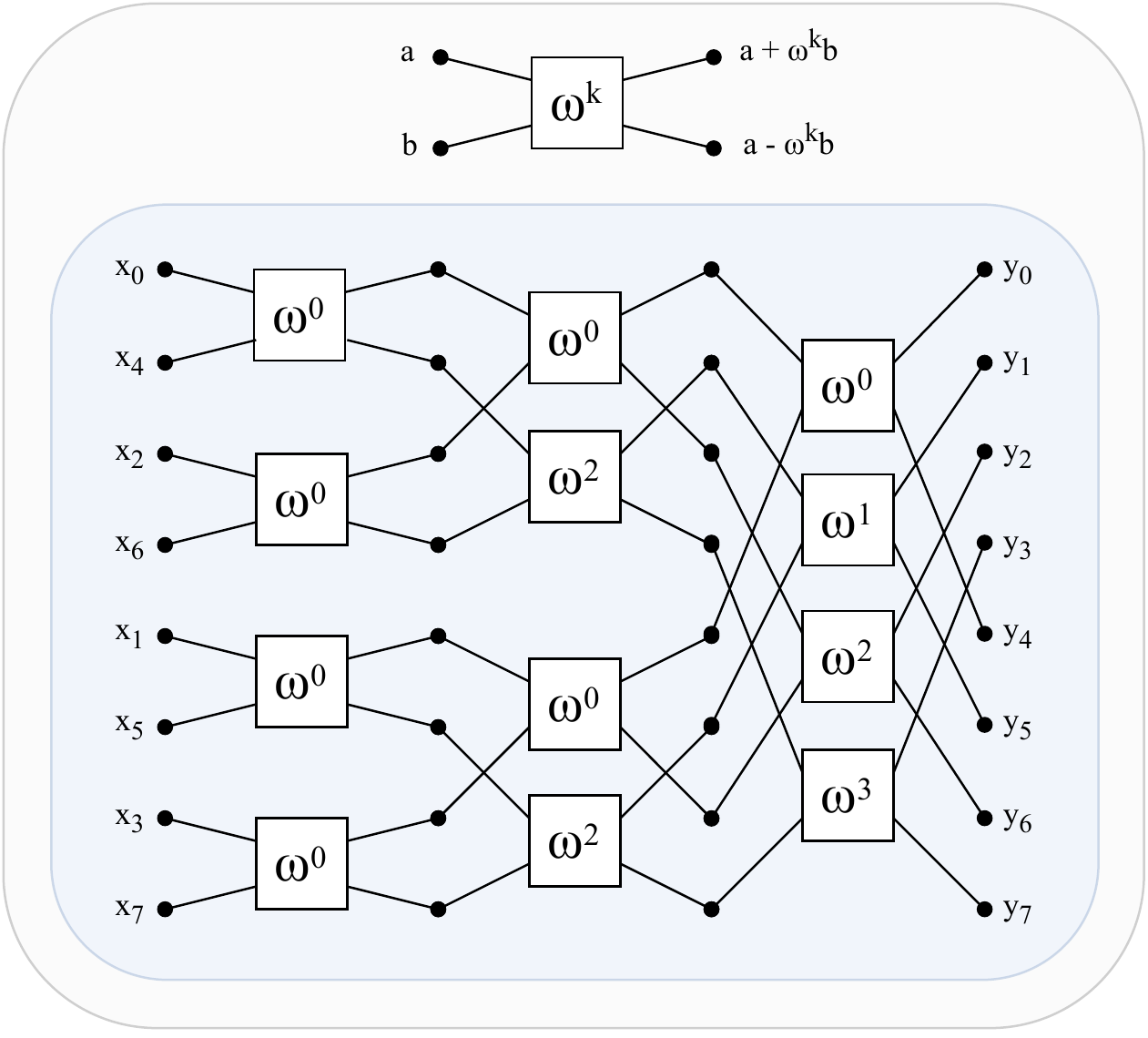}
    \caption{Diagram of the Cooley-Tukey algorithm \cite{cooley1965algorithm}, performing the classical FFT on an input $x\in\R^n$. Each white box performs an elementary radix-2 operation with the root of unity $\omega = e^{i2\pi/n}$.}
    \label{fig:classical_fft}
\end{figure}

The classical algorithm for performing Fast Fourier Transform (FFT) uses a butterfly-shaped diagram \cite{cooley1965algorithm}, shown in Fig.\ref{fig:classical_fft}. Our goal is to inspire from the classical FFT diagram and perform the same operation with quantum circuits. Namely, the unitary matrix, once restricted to the unary basis, must implement the FFT matrix.  
With an input $x \in \R^n$, The FFT matrix $F_n$ is given by:

\begin{equation}
    F_n = \begin{pmatrix}
        1&1&1&\cdots&1 \\
        1&\omega & \omega^2 & \cdots & \omega^{(n-1)} \\
        1&\omega^2 & \omega^4 & \cdots & \omega^{2(n-1)} \\
        \vdots & \vdots & \vdots & \ddots & \vdots \\
        1&\omega^{n-1} & \omega^{2n-2} &\cdots & \omega^{(n-1)^2} \\
    \end{pmatrix} 
\end{equation}

where $\omega^k = e^{i\frac{2\pi}{n}k}$ is the $n^{th}$-root of unity raised to some power $k$ based on the position of the radix. Note that $F_n$ is not unitary, but its scaled version $F_n/\sqrt{n}$, is unitary. Therefore, we will implement the scaled version using our quantum circuit. As shown in Fig.\ref{fig:classical_fft}, the input $x$ is permuted, which we will have to take into account when loading data quantumly.

The classical FFT algorithm is decomposed into several radix-2 operations (Fig.\ref{fig:classical_fft}). Each one itself is a matrix multiplication which transforms the input $[a, b]$ into $[a+\omega^k b, a-\omega^k b]$. In matrix multiplication terms, each of these operations applies the matrix
$\begin{pmatrix}
    1&\omega^k\\
    1&-\omega^k\\
\end{pmatrix}$.

\begin{figure}
    \centering
    \includegraphics[width=\linewidth]{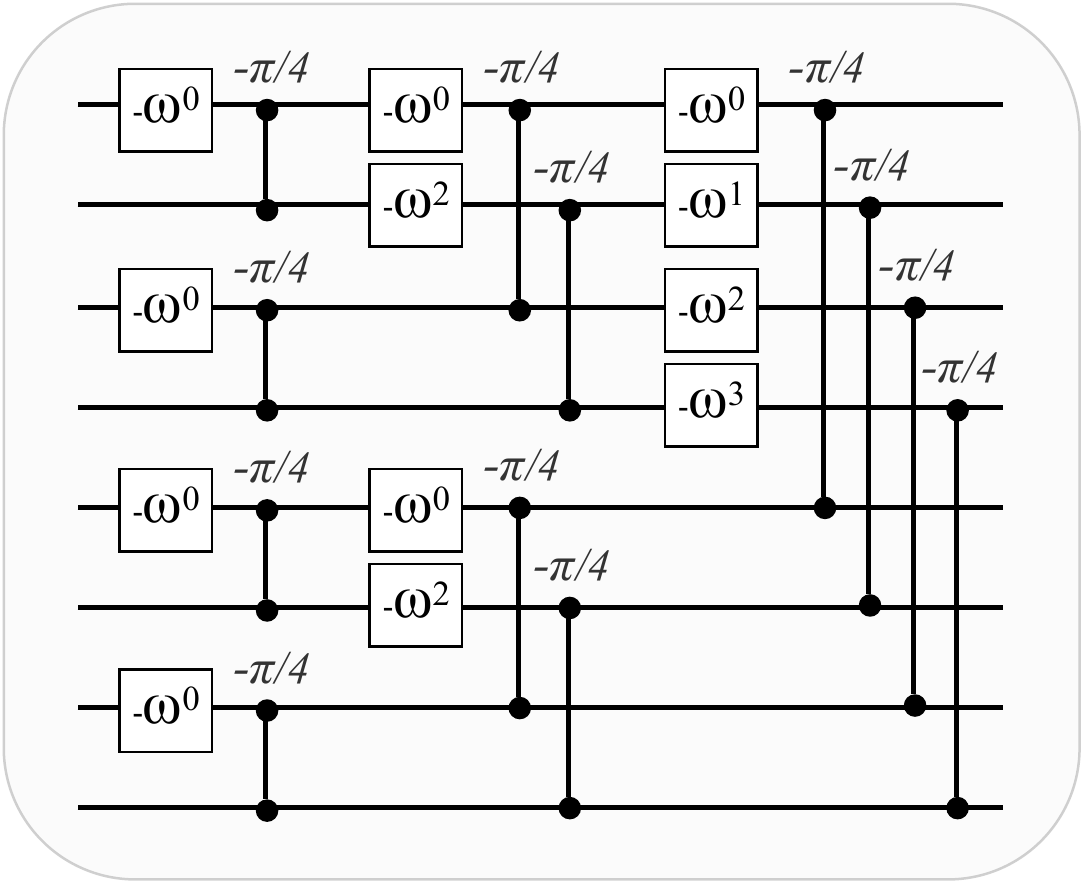}
    \caption{Quantum circuit for implementing a Fourier Transform on the unary basis. Single qubit gates are phase gates. Vertical lines are $RBS$ gates with $-\pi/4$ angle. It reproduces the classical FFT butterfly circuit (Fig.\ref{fig:classical_fft}) by replacing each radix-2 operation with the phase gate followed by the RBS gate. The input must be a vector $\ket{x}$ encoded in the unary basis, with the right permutation. The output will be $\ket{\hat{x}}$.}
    \label{fig:Unary-QFT}
\end{figure}

Quantumly, we want to reproduce the idea of the classical FFT diagram. It results in the circuit shown in Fig.\ref{fig:Unary-QFT}. To reproduce the action of the radix-2 transforms (scaled by a factor of $1/\sqrt{2}$ to make it unitary) on two qubits, we use one single qubit gate and one $RBS$ gate (Eq.\ref{eq:RBS}). The single qubit gate is a phase gate 
$\begin{pmatrix}
    1&0 \\
    0&-\omega^k \\
\end{pmatrix}$
applied on the top qubit only. Then, we apply an $RBS$ gate with an angle of $-\pi/4$. Overall, this applies the following unitary:

\begin{equation}
\begin{aligned}
    &\begin{pmatrix}
        1&0&0&0 \\
        0&\frac{1}{\sqrt{2}}&-\frac{1}{\sqrt{2}}&0 \\
        0&\frac{1}{\sqrt{2}}&\frac{1}{\sqrt{2}}&0 \\
        0&0&0&1 \\
    \end{pmatrix} 
    \cdot
    \begin{pmatrix}
        1&0&0&0 \\
        0&1&0&0 \\
        0&0&-\omega^k &0 \\
        0&0&0&-\omega^k \\      
    \end{pmatrix} \\
    &=
    \begin{pmatrix}
        1&0&0&0 \\
        0&\frac{1}{\sqrt{2}}&\frac{\omega^k}{\sqrt{2}}&0 \\
        0&\frac{1}{\sqrt{2}}&-\frac{\omega^k}{\sqrt{2}}&0 \\
        0&0&0&-\omega^k \\
    \end{pmatrix} 
    \end{aligned}
\end{equation}

The  above unitary, once restricted to the unary basis, namely $\{\ket{01},\ket{10}\}$ (middle rows and columns) in the case of two qubits, is exactly the desired radix operation :
$
    \frac{1}{\sqrt{2}}\begin{pmatrix}
        1&\omega^k\\
        1&-\omega^k\\
    \end{pmatrix}
$

We apply these operations exactly in the manner of the classical FFT architecture. For an $n$-dimensional vector, we would thus require $n$ qubits, $\log(n)$ depth and $n\log(n)$ gates. 

The Unary-QFT is meant to be applied on a unary quantum state, after a vector data loader for instance. Note that the input to the circuit needs to be the permuted version of the vector. This is a fixed permutation, and we can construct our data loader accordingly.

Finally, it is simple to apply the inverse Fourier transform in a similar way, named Unary-IQFT. It is simply the inverse of the above circuit (Fig.\ref{fig:Unary-QFT}), along with the right permutation of the input data. 

In the rest of this work, we will use the following equations. Given a normalised real vector $x = (x_1,\cdots,x_M)$, and its Fourier Transform $\hat{x} = (\hat{x}_1,\cdots,\hat{x}_M)$, the $QFT$ operation in unary basis and its inverse, the $IQFT$ operation, can be defined as follows:
\begin{equation}
\label{eq:QFT_equations}
\begin{aligned}
    &QFT (\sum_i x_i \ket{e_i}) = \sum_i \hat{x}_i \ket{e_i} \quad \text{and} \\ &IQFT (\sum_i \hat{x}_i \ket{e_i}) = \sum_i x_i \ket{e_i}
    \end{aligned}
\end{equation}


\subsection{Quantum Circuits as Trainable Linear Transforms}\label{sec:TrainableQuantumCircuits}

To mimic the learnable part of the Fourier Layer, we need to perform quantumly some sort of matrix multiplication in the unary basis. 
With this in mind, we propose to use \emph{Quantum Orthogonal Layers} \cite{landman2022quantum}, namely parameterised quantum circuits using hamming-weight preserving gates only. 
In our case, we are consequently ensured to preserve the superposition in the unary basis, before applying the final Inverse Fourier Transform. 
The unitary considered have real values and once restricted to the unary basis, correspond to orthogonal matrices, hence their name. 
In the context of the Fourier Layer, we are now able to apply a learnable linear transform (see Section \ref{sec:classical_fno}), by implementing orthogonal matrix multiplications with trainable parameters. 

Several circuits are possible, but we will mostly use the \emph{Butterfly} circuit shown in Fig.\ref{fig:butterfly}: it has the same layout as the previous Unary-QFT (Section \ref{sec:unaryQFT}), and for this reason has a logarithmic depth if the hardware connectivity allows it. For an input $x\in\R^n$, the Butterfly quantum circuit has $O(n\log(n))$ parameterised gates. The corresponding matrices, once restricted to the unary basis, lie in a subgroup of orthogonal matrices. 
Other circuits, with nearest-neighbours qubits connectivity, are also usable in our case. They differ in their expressivity, number of parameters, and depth. 

In the next sections, we will detail how to apply such circuits in a controlled fashion. This will become useful as $K$ such orthogonal matrix multiplications will be applied on the first $K$ \emph{modes} (see Section \ref{sec:classical_fno}), after the Fourier Transform.

\begin{figure}
    \centering
    \includegraphics[width=0.8\linewidth]{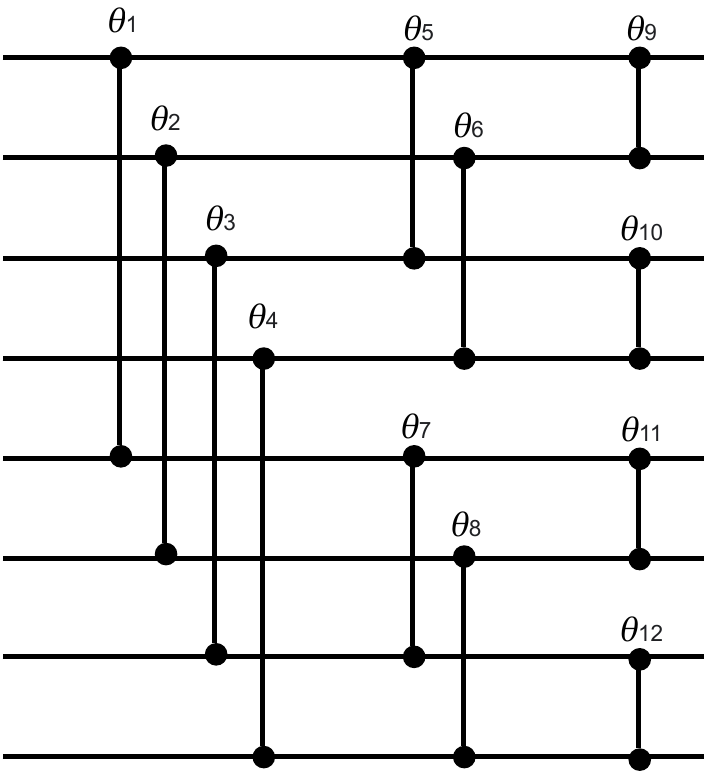}
    \label{fig:butterfly}    
    \caption{parameterised quantum circuits with a butterfly shape will take the role of the linear transforms (matrix multiplications) in the Fourier Layer. Given complete hardware connectivity, their depth is logarithmic in the number of qubits.}
    \label{fig:butterfly}
\end{figure}

\section{Quantum Circuits for Fourier Layer}\label{sec:quantumFourierLayer}

Using the circuits from the previous section as building blocks, we propose three quantum circuits for classical Fourier Layer (FL), named Sequential (Section \ref{sec:sequential_qfno}), Parallel (Section \ref{sec:parallel_qfno}), and Composite (Section \ref{sec:compound_qfno}) Quantum Fourier Layer (QFL) respectively. These quantum circuits are differentiated based on how the intermediate matrix multiplications in the classical FL are implemented using the learnable quantum circuit.
We compare their computational complexities (see Table \ref{tab:complexity}) and their efficiency in practice in the following sections.

To reproduce the classical Fourier Layer (FL) (see Section \ref{sec:classical_fno}) using quantum circuits, we need the final quantum state to correspond to the result of the classical FL, as shown in Eq.\ref{eq:classical_out}. Therefore, we expect ideally a quantum output state of the following form:


\begin{equation}
\begin{aligned}
        \ket{y} &= \sum_i \ket{y_i}\ket{e_i}  =\sum_i^{N_c}\sum_{j}^{N_s}y_{ij}\ket{e_{j}}\ket{e_{i}} \quad \text{where} \\
        y_{ij} &= IFT\left(\left[(\hat{b}_{ij})_{j\in[1,K]} ,  (a_{ij})_{j\in[K+1,N_s]}\right]\right)_{j}
        \end{aligned}
    \label{eq:ideal_quantum_output_y}
\end{equation}

which simply means that the matrix $y$ is encoded in the unary basis, as is Eq.\ref{eq:quantum_matrix_data_state}, and $y_{ij}$ is the $j^{th}$ component of the resulting vector $y_{i}$ defined in Eq.\ref{eq:classical_out}. Note that there are no normalisation factors in the above equation, as the rows of the matrix $(a_1,\cdots, a_{N_c})$ are assumed normalised, and the normalisation is preserved through the circuit.


For our three quantum circuit proposals— the Sequential circuit (\ref{sec:sequential_qfno}), the Parallel circuit (\ref{sec:parallel_qfno}), and the Composite circuit (\ref{sec:compound_qfno})—we will compare their output quantum state to the desired one (Eq.\ref{eq:ideal_quantum_output_y}), showing how well we are able to replicate the classical operation. 

Our conclusions are the following: the Sequential circuit is returning the desired output $\ket{y}$ but might have a prohibitive depth for near-term application. On the other hand, both Parallel and Composite circuits are designed to have a shorter depth, at the cost of producing a slightly different quantum output. Interpreting these alternative outputs is complicated, and numerical simulations in Section \ref{sec:Experiments} will assess their quality. 

\begin{figure*}[!htb]
    \centering
    \includegraphics[width=\linewidth]{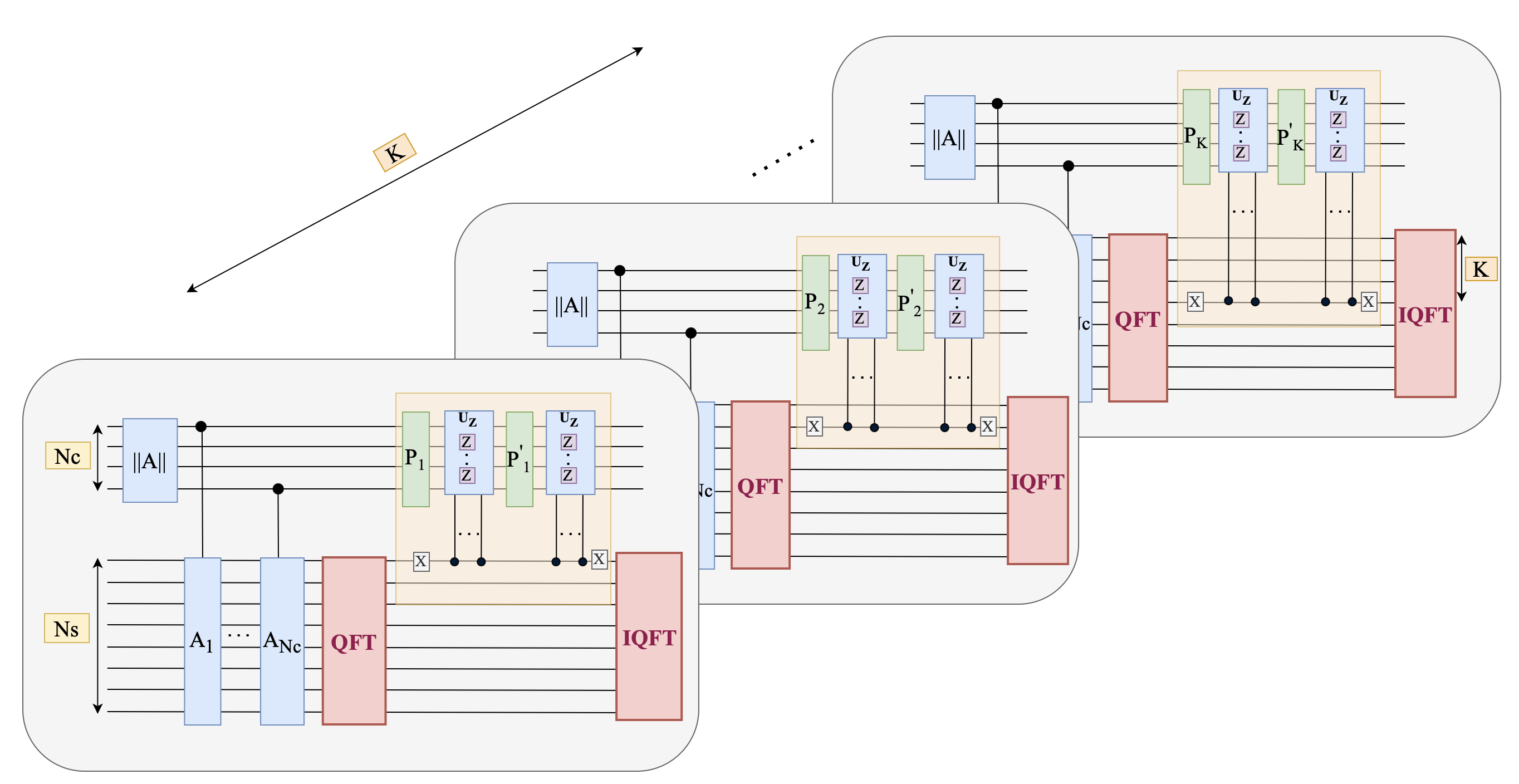}
    \caption{\textbf{Parallel QFL:} Parallelised version of the Sequential Quantum Circuit to minimize the depth of the learning part, thus making it more efficient when deployed on noisy hardware. For each mode (out of the top $K$) in the transformed input, there is a different circuit to perform the parameterised matrix transform.}
    \label{fig:parallel_qfno_circuit}
\end{figure*}

The three proposals are closely related, as shown in Figures \ref{fig:sequential_qfno_circuit}, \ref{fig:parallel_qfno_circuit}, and \ref{fig:compound_qfno_circuit}. All circuits start with $\ket{0}^{N_c}\otimes \ket{0}^{N_s}$, where the top and lower registers are used respectively to encode the $N_c$ rows and the $N_s$ columns of the input matrix $A$. Indeed, all circuits start by loading the input matrix $A$ as explained in Section \ref{sec:dataloader}. We recall that $N_s$ is the number of samples that are used to encode each initial condition function of the PDE, as a vector. $N_c$ is another dimension we use to extend this vector as a matrix. We usually have $N_s \gg N_c$. After this step, the Unary-QFT from Section \ref{sec:unaryQFT} is applied on the lower register. At the very end, the inverse QFT is applied similarly, followed by a measurement of both registers. Between the two Fourier Transforms lies the core difference between the three proposals: the implementation of the $K$ matrix multiplications (see Fig.\ref{fig:classical_FNO_fig1} and Section \ref{sec:classical_fno}). This part is a trade-off between circuit depth, circuit repetitions, and correspondence with the classical Fourier Layer. 


\subsection{Sequential Circuit for Fourier Layer} 
\label{sec:sequential_qfno}


\begin{figure*}[!htb]
    \centering    \includegraphics[width=\linewidth]{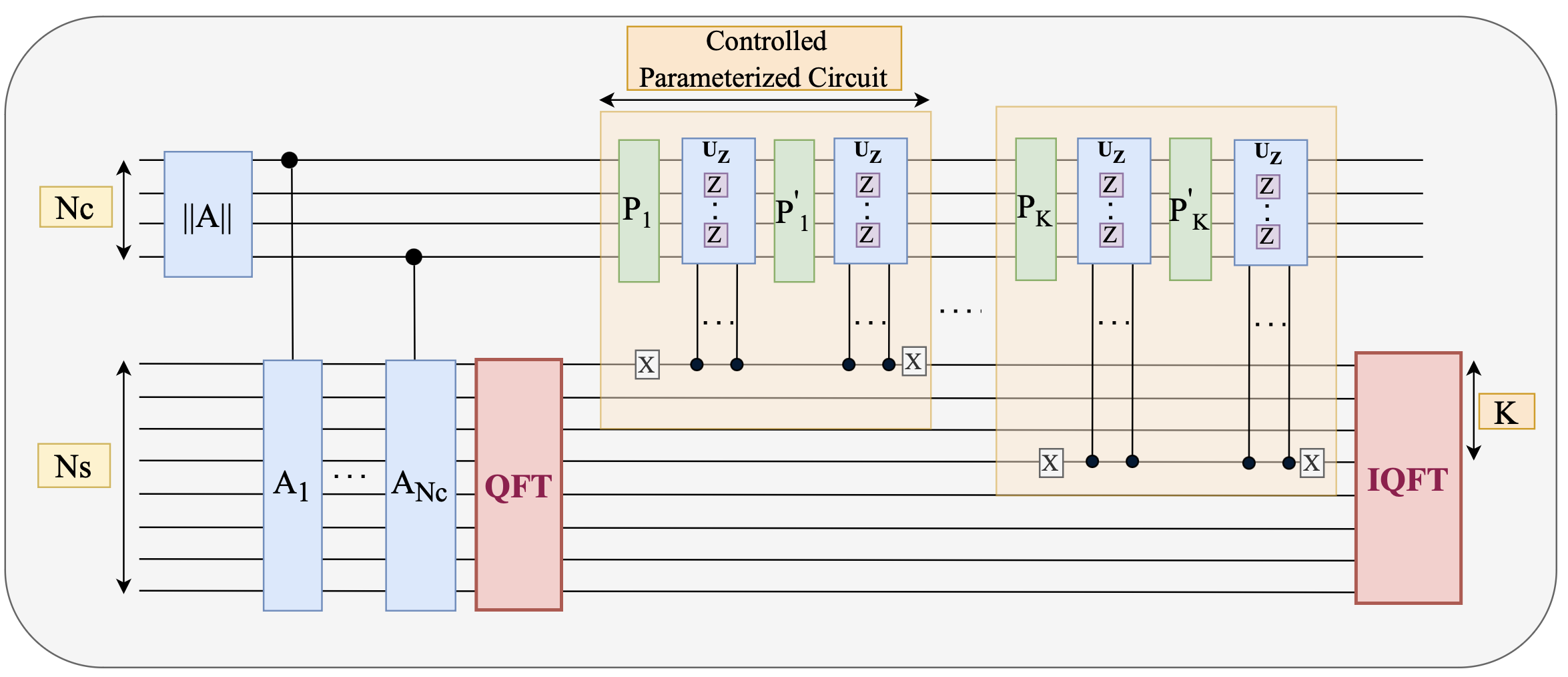}
    \caption{
    \textbf{Sequential-QFL:} The proposed Sequential Quantum Circuit, which replicates the classical FNO operation from {Fig.\ref{fig:classical_FNO_fig1} (if we measure at the end}). Further details regarding it are given in Section \ref{sec:sequential_qfno}. The yellow box comprises a controlled parameterised circuit having $j^{th}$ qubit ($1\leq j \leq K$) of the lower register as the controlling qubit.}
\label{fig:sequential_qfno_circuit}
\end{figure*}





This first proposal is represented as Fig.\ref{fig:sequential_qfno_circuit}.
As explained, the Sequential circuit starts by loading the input matrix $A$, which in general is itself the output of the previous Fourier Layer. Therefore, the first part of the circuit is nothing else than the circuit displayed in Fig.\ref{fig:matrixdataloader}. Its depth is $O(\log(N_c)+2N_c\log(N_s))$. The resulting state is:

\begin{equation}
    \sum_{i}^{N_c}\sum_{j}^{N_s} a_{ij}\ket{e_{i}}\ket{e_{j}}
    \label{circuit_1_before_qft}
\end{equation}

To follow the classical operation, the Fourier Transform should be applied to each of the rows. Thus, we apply the Unary-QFT operation on the lower register, currently encoding a superposition of these row vectors. 


We first rewrite the previous state as:

\begin{equation}
    \sum_{i}^{N_c}\ket{e_{i}}(\sum_{j}^{N_s} a_{ij}\ket{e_{j}})
    \label{circuit_1_before_qft}
\end{equation}

And then apply the $QFT$ operation, as in Eq.\ref{eq:QFT_equations}, to the lower register:
\begin{equation}
\begin{gathered}
\sum_{i}^{N_c}\ket{e_{i}}QFT(\sum_{j}^{N_s} a_{ij}\ket{e_{j}}) \\
=\sum_{i}^{N_c}\ket{e_{i}}\sum_{j}^{N_s} \hat{a}_{ij}\ket{e_{j}}
    = \sum_{i}^{N_c}\sum_{j}^{N_s} \hat{a}_{ij}\ket{e_{i}}\ket{e_{j}}
    \label{eq:1_after_qft}
\end{gathered}
\end{equation}
where $\hat{a}_{ij}$ corresponds to the classical, row-wise, Fourier Transform of $A$. Note that the resulting state has kept its superposition in the unary basis. \\


Next, we realize the learnable linear transform, namely $K$ matrix multiplications, as in the middle of the classical Fourier Layer (Fig.\ref{fig:classical_FNO_fig1}). Each matrix $W^k$, to follow the classical notations from Section \ref{sec:classical_fno}, will now become an orthogonal matrix in the quantum case.
This multiplication is done with the parameterised quantum circuits from Section \ref{sec:TrainableQuantumCircuits}, which preserves the unary basis. We choose to use the learnable butterfly circuit from Fig.\ref{fig:butterfly}, for its shallow depth. As explained before, applying these learnable circuits effectively applies an orthogonal matrix multiplication in the unary basis.
The main challenge, which will differentiate the three proposals, consists in applying these matrix multiplications to the first $K$ columns $(\hat{a}^1,\cdots,\hat{a}^K)$, independently. Contrary to the previous QFT, the linear transform should act on the columns, encoded in the upper register. 



As shown in Fig.\ref{fig:sequential_qfno_circuit}, we propose to apply sequentially $K$ such parameterised circuits $P_1, \cdots, P_K$ (butterfly circuits) on the top register.
{To ensure independent transformations are applied on each column $\hat{a}^j$, we propose a controlled implementation of the parameterised circuit, or, a \textit{Controlled parameterised Circuit}. It applies the parameterised circuit to the upper register only when a particular qubit of the lower register is not in the activated state ($\ket{1}$). For the circuit $P_j$, controlled by $j^{th}$ qubit of the lower register, this controlled implementation begins with the application of circuit $P_j$ on the upper register, followed by the application of Controlled Z-gates to certain qubits in the upper register if the $j^{th}$ qubit in the lower register is in state $\ket{0}$. 
This can be seen as a controlled implementation of some matrix $U_Z$ (see Fig.\ref{fig:sequential_qfno_circuit}) and thus, we denote this transformation by $CU_Z$. \\
\noindent
The set of qubit indices for applying these Z-gates are selected using the following rule: \textit{each RBS-gate of the parameterised circuit $P_j$ has a $Z$-gate applied to exactly one of its two qubits.} For instance, if $P_j$ has nearest neighbour connectivity, then the desired set can be the set of even or odd qubit indices.
Subsequently, we apply another parameterised circuit, similar to $P_j$, on the upper register, but this time reversing the order in which the RBS gates are applied, denoting this circuit by $P_j^{\prime}$. Finally, we repeat the controlled Z-gate operations on the same qubits. This completes the implementation of the controlled parameterised circuit. Let us now see how this implementation achieves the desired task. \\
Fig. \ref{fig:circuit_z_mirror} shows the controlled version of the parameterised circuit from  Figure \ref{fig:butterfly}. Here, the lowest qubit controls all the Z-gates and thus, is the controlling qubit for the parameterised circuit. The complete circuit includes the parameterised circuit ($P_j$) followed by controlled Z-gates on some of the qubits ($CU_Z$). These qubits are selected using the following criterion: \textit{Every RBS-gate in the circuit has exactly one of its two qubits lying in this set.}
Finally, the circuit $P_j$ is applied again but this time the order of RBS-gate is reversed ($P_j^{\prime}$). Denoting the angles $[\theta_1, \theta_2,...., \theta_{M}]$ by $\mathbf{\theta}$, where $M$ is the total number of parameterised RBS-gates in $P_j$, and writing the circuit $P_j$ as $P_j({\theta})$, the following relation holds between $P_j$ and $P_j^{\prime}$:
\begin{equation}
    P_j^{\prime}(\theta) = P_j^{\dagger}(-\theta)
\end{equation}
where $P_j^{\dagger}$ denotes the conjugate transpose of $P_j$.

Before proceeding further on the implementation of the controlled parameterised circuit, we first discuss a claim below regarding the existence of $U_Z$ for any possible parameterised butterfly circuit.

\vspace{2.00mm}
\noindent
\textbf{Claim 4.1}\textit{ For an N-qubit butterfly-shaped parameterised circuit described in Section \ref{sec:quantumtools}, with N=$2^a$ for any whole number a, there exists a set of qubit indices $\mathcal{I}_a$ such that every RBS-gate in the circuit has exactly one of its qubits' index lying in this set.} \\
\textbf{Proof:} We show this proof 
by recurrence. We first establish the base case for a=1 (or N=2). For this we know $\mathcal{I}_1$ will be either \{1\} or \{2\}. \\
Now, it is given that $\mathcal{I}_a$ exists for some $a>1$. Since $\mathcal{I}_a$ consists of exactly one of the qubit index for every RBS-gate, then indices not in $\mathcal{I}_a$ (denoted by $\mathcal{I}^{'}_a$) cannot have any RBS-gate between them. Also, since exactly one of every RBS-gate indices is in $\mathcal{I}_a$, therefore the other index lies in $\mathcal{I}^{'}_a$. Thus, $\mathcal{I}^{'}_a$ is also a solution set for this problem.  
Now, if we combine the two $N=2^a$ qubit circuits (corresponding to $\mathcal{I}^1_a$ and $\mathcal{I}^2_a$), then we see that there is no new connection on the index set $\mathcal{I}^{'1}_aU\mathcal{I}^{2}_a$. Thus, this can correspond to the set $\mathcal{I}_{a+1}$. Hence, we proved that if $\mathcal{I}_a$ exists, then $\mathcal{I}_{a+1}$ also exists.

\vspace{2.00mm}
\noindent
 Going further, we now use Identity \ref{identity_rbs} for RBS gates to arrive at the following equation:
\begin{equation}
    U_ZP_j(\theta) = P_j(-\theta)U_Z
\end{equation}
Using this, we arrive at the following relation:
\begin{equation}
\begin{aligned}
    P_j^{\prime}(\theta)U_ZP_j(\theta) &= P_j^{\dagger}(-\theta)U_ZP_j(\theta) \\
    &= P_j^{\dagger}(-\theta)P_j(-\theta)U_Z \\
    &= U_Z
\end{aligned}
\label{eq:p_u_p}
\end{equation}

\begin{figure}[!htb]
    \centering
    \includegraphics[width=\linewidth]{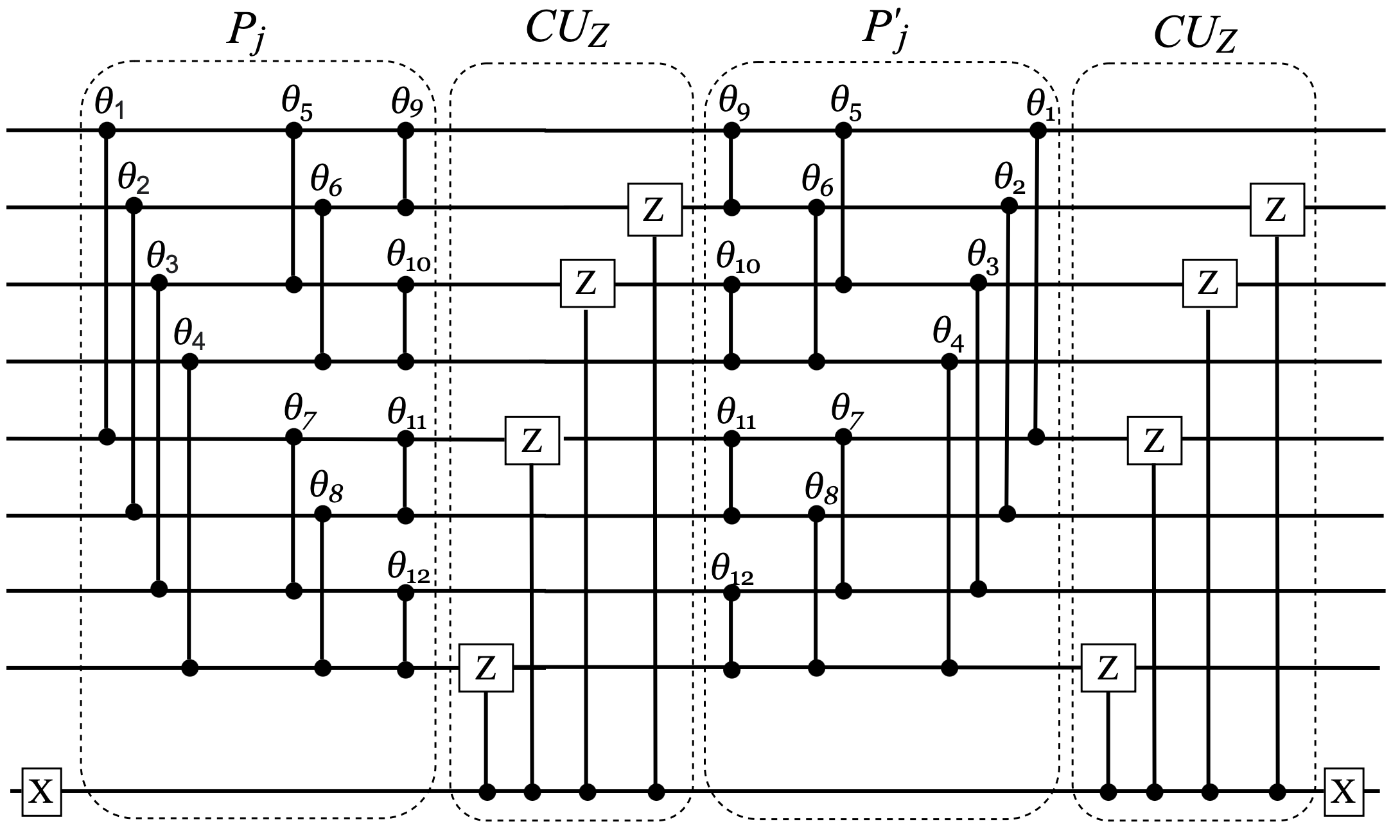}
    \caption{parameterised Butterfly Circuit followed by Z gates on certain qubits and its flipped version around the vertical axis. The dash-enclosed region shows the last layer RBS gates of the parameterised circuit being cancelled out by first layer RBS gates in its flipped version, using the RBS identities in figure \ref{fig:rbs_identity}. Similarly for the second last layer after the last layer is cancelled out and so on. Thus, eventually, only Z gates will remain.}
    \label{fig:circuit_z_mirror}
\end{figure}

Thus, it can be observed that when the lowest qubit in Figure \ref{fig:circuit_z_mirror} is in state $\ket{0}$, each RBS gate in the last layer of $P_K$ is being cancelled out by the RBS gate in the first layer of $P_K^{\prime}$. After this, the second last layer operations get cancelled by the second layer of $P_K^{\prime}$ and finally only Z-gate operations ($U_Z$) will remain. 
Finally, the application of another $CU_Z$ unitary results in $U_Z$ being re-applied to the upper register and using $U_ZU_Z=I$, we can conclude that the state of the upper register is preserved. Thus, applying X-gate on the lowermost qubit again retains the initial state before the application of this circuit.\\

On the other hand, when the lowest qubit is in state $\ket{1}$, no Z-gates are applied, and the initial state of the remaining qubits is transformed by $P_K$ and $P_K^{\prime}$.
This corresponds to a controlled version of a parameterised circuit, namely $P_j P'_j$. Note that it is slightly different from applying a controlled $P_j$ only, whose implementation remains an open question.\\

Let us now apply the Controlled version of $P_1$ on the state in Eq.(\ref{eq:1_after_qft}).
After re-arranging terms and applying $P_1$, it can be written as:
\begin{equation}
\sum_j^{N_s}\ket{e_j}P_1(\sum_i^{N_c}\hat{a}_{ij}\ket{e_i})
\end{equation}


Now, on applying the $CU_Z$ unitary using the first qubit of the lower register as the control (applied when this qubit is in state $\ket{0}$), the state becomes:
\begin{equation}
\ket{e_1}P_1(\sum_i^{N_c}\hat{a}_{ij}\ket{e_i}) + \sum_{j\neq 1}^{N_s}\ket{e_j}U_Z P_1(\sum_i^{N_c}\hat{a}_{ij}\ket{e_i})
\end{equation}
Now, applying the flipped circuit $P^{'}_1$ transforms the state to:
\begin{equation}
\ket{e_1}P^{'}_1 P_1(\sum_i^{N_c}\hat{a}_{ij}\ket{e_i}) + \sum_{j\neq 1}^{N_s}\ket{e_j}P^{'}_1U_Z P_1(\sum_i^{N_c}\hat{a}_{ij}\ket{e_i})
\label{eq:after_mirror}
\end{equation}

Using $P^{'}_1U_Z P_1 = U_Z$  from Eq.(\ref{eq:p_u_p}), the state reduces to:
\begin{equation}
\ket{e_1}P^{'}_1 P_1(\sum_i^{N_c}\hat{a}_{ij}\ket{e_i}) + \sum_{j\neq 1}^{N_s}\ket{e_j}U_Z(\sum_i^{N_c}\hat{a}_{ij}\ket{e_i})
\label{eq:after_mirror_updated}
\end{equation}
Finally, repeating the controlled application of $U_Z$ leads to:
\begin{equation}
\ket{e_1}P^{'}_1 P_1(\sum_i^{N_c}\hat{a}_{ij}\ket{e_i}) + \sum_{j\neq 1}^{N_s}\ket{e_j}(\sum_i^{N_c}\hat{a}_{ij}\ket{e_i})
\label{eq:after_mirror_updated_latest}
\end{equation}
}

Applying K such controlled circuits, where the $j^{th}$ circuit on the upper register is controlled by $j^{th}$ qubit in the lower register, and re-arranging the terms in the last equation leads to the following state: 
\begin{equation}
\sum_{j=1}^KP^{'}_j P_j(\sum_i^{N_c}\hat{a}_{ij}\ket{e_i})\ket{e_j} + \sum_{j=K+1}^{N_s}(\sum_i^{N_c}\hat{a}_{ij}\ket{e_i})\ket{e_j}
\label{eq:after_K_butteflies}
\end{equation}

We now want to compare the state obtained in Eq.(\ref{eq:after_K_butteflies}) to the classical output of the FNO from Eq.(\ref{eq:classical_before_iqft}), before the final inverse Fourier Transform. We recall that in the classical case, each of the first $K$ columns $\hat{a}^j$ was multiplied by an independent matrix $W^j$. In the quantum case, we need to understand if the same operation is applied, and with which matrix $W_Q^j$.

For all $j \in [1,K]$, we saw in Eq.(\ref{eq:after_K_butteflies}) that $P'_jP_j$ was applied to the unary encoding of $\hat{a}^j$. We denote respectively by $W_{P_j}$ and $W_{P'_j}$ the unary matrix of $P_j$ and $P'_j$. Each matrix is the $N_c\times N_c$ submatrix of the whole unitary of size $2^{N_c}\times 2^{N_c}$, corresponding to the basis states of unary vectors (see Section \ref{sec:TrainableQuantumCircuits}). Therefore, considering only the top register, the $j^{th}$ operation {$P^{'}_jP_{j}$ corresponds to the sub-matrix $W_Q^{j}$:
\begin{equation}
    W^j_Q = W_{P_j^\prime}W_{P_j} 
\end{equation}
This matrix $W^j_Q$ is the quantum implementation corresponding to the matrix $W_j$ used in the implementation of classical FNO (eq. \ref{eq:classical_before_iqft}).
The overall matrix $(\hat{a}_{ij})$ can be decomposed into $N_s$ vectors $\hat{a}^j = (\hat{a}_{ij})_{i\in[1,N_c]}$. Then, for the first $K$ vectors $\hat{a}^j \in \mathbb{R}^{N_c}$ we will have $\hat{b}^j = W_Q^j \hat{a}^j$ 
, where this $\hat{b}^j$ is the quantum counterpart for one used in classical FNO. Thus, the state after these controlled parameterised circuits can be written as:
\begin{equation}
\sum_i^{N_c}\left(\sum_j^K\hat{b}_{ij}\ket{e_i}\ket{e_j} + \sum_{j=K+1}^{N_s}\hat{a}_{ij}\ket{e_i}\ket{e_j}\right)
\label{eq:final_after_K_butteflies}
\end{equation}}
Finally, the output state of this circuit after $IQFT$ on the lower register becomes: 

\begin{equation}
   \sum_{i}^{N_c}\ket{e_{i}}IQFT\left (\sum_{j}^{K} \hat{b}_{ij}\ket{e_{j}} + \sum_{j=K+1}^{N_s} \hat{a}_{ij}\ket{e_{j}}\right) 
    \label{eq:final_qc_1}
\end{equation}

Since $IQFT(\sum_{i}\hat{x}_i\ket{e_{i}})$ = $\sum_{i}x_i\ket{e_{i}}$, where $IFT(\hat{x})=x$, this implies that $j^{th}$ component of $IFT$ would be same as the coefficient of $j^{th}$ state in $IQFT$. From this, we can conclude that the state in Eq.(\ref{eq:ideal_quantum_output_y}) is equivalent to the state in Eq.(\ref{eq:final_qc_1}) and thus, this circuit replicates the classical operation. Finally, let us now discuss the depth complexity of this circuit.
\\

\noindent

\textbf{Depth Complexity ($d$).} 
Based on the discussion of the Sequential QFL circuit, it can be divided into four parts : (a) unary loading of the input matrix ($d_{load}$), (b) applying $QFT$ on the lower register ($d_{qft}$), (c) applying $K$ Controlled Parameterised Circuits on the upper register ($d_{cpc}$) and (d) applying  inverse $QFT$ on the lower register ($d_{iqft}$). Thus, the depth of the complete Sequential QFL circuit becomes:

\begin{equation}
    \begin{aligned}
        =& d_{load} + d_{qft} + Kd_{cpc} + d_{iqft} \\
        =& \left(N_c\text{log}(N_s)+\text{log}(N_c) \right) + \text{log}(N_s) + \\
        &K(N_c+2\text{log}(N_c)) + \text{log}(N_s) \\
        =& (N_c+2)\text{log}(N_s) + (2K+1)\text{log}(N_c) + KN_c \\
    \end{aligned}
    \label{eq:depth_seq}
\end{equation}


\subsection{Parallelised Circuit for Fourier Layer} 
\label{sec:parallel_qfno}

For the Sequential QFL discussed in the previous subsection, the depth complexity of the learnable part is linear in the number of modes (K). Given the multiplicative noise model for NISQ devices, this linear dependence might hinder learning. A helpful modification then can be parallelising the learnable butterfly circuits, which can make the learning in the presence of noise more efficient and reduce the circuit's depth complexity.
Fig.\ref{fig:compound_qfno_circuit} shows this modified version of the Sequential QFL, consisting of $K$ quantum circuits operating in parallel and each implementing only one learnable circuit controlled by one of the top $K$ qubits in the lower register. 
As all the circuits up to the learnable part are similar to the sequential circuit, we can directly write the state after the $QFT$ using Eq.( \ref{eq:1_after_qft}), as:
\begin{equation}
\left[\sum_{j}^{N_s}\sum_{i}^{N_c}(\hat{a}_{ij})_{k}\ket{e_{i}}_{k}\ket{e_{j}}_{k}\right]_{k=1}^{K}
\end{equation}
where the index $k$ denotes the $k^{th}$ parallel circuit and $(\hat{a}_{ij})_k$, $\ket{e_i}_k$ denote the coefficient $\hat{a}_{ij}$, state $\ket{e_i}$ corresponding to this $k^{th}$ circuit respectively. Also, 
in the $k^{th}$ parallel circuit, the learnable butterfly part is controlled by the $k^{th}$ qubit of the lower register. 
We recall that the parameterised circuit applied on the top register is effectively mapping the vector $\hat{a}^j$ to $\hat{b}^j$ (see Eq.(\ref{eq:final_qc_1})) and thus, we can write the updated state of the circuits as: 
\begin{equation}
\left[\sum_{j\neq k}\sum_{i}^{N_c}(\hat{a}_{ij})_{k}\ket{e_{i}}_{k}\ket{e_{j}}_{k}+\sum_{i}^{N_c}(\hat{b}_{ij})_k\ket{e_{i}}_{k}\ket{e_{k}}_{k}\right]_{k=1}^{K}
\label{eq:3_before_iqft}
\end{equation}
Now, applying $IQFT$ on the lower register in each of the circuits independently:
\begin{equation}
\begin{aligned}
  &\left[ \sum_{i}^{N_c}\ket{e_{i}}_{k}
  IQFT
  \left((\hat{b}_{ik})_{k}\ket{e_{k}}_{k} 
  + \sum_{j\neq k}(\hat{a}_{ij})_{k}\ket{e_{j}}_{k}
  \right)
  \right]_{k=1}^{K} 
  \label{eq:final_state_eq_parallel}
\end{aligned}
\end{equation}
We denote coefficients of this state corresponding to $k^{th}$ circuit by $(c_{ij})_k$ and thus re-writing it as:
\begin{equation}
\begin{aligned}
  \left[ 
  \sum_{i}^{N_c}
  \sum_{j}^{N_s}
  (c_{ij})_k\ket{e_{i}}_{k}\ket{e_{j}}_{k}
  \right]_{k=1}^{K} 
  \label{eq:c_state_circuit_2}
\end{aligned}
\end{equation}
where all of the $(c_{ij})_k$ are explicitly given by using the equation for the Fourier transform as follows:
\begin{equation}
    (c_{ij})_k =\frac{1}{N_s}\left ( \sum_{j\neq k} (\hat{a}_{ij})_ke^{i \frac{2\pi j}{N_s}} +(\hat{b}_{ik})_ke^{i \frac{2\pi k}{N_s}} \right)
    \label{eq:c_circuit_2}
\end{equation}
Similarly, writing $c_{ij}$ for the sequential circuit discussed in the paper (using eq. \ref{eq:final_qc_1}):
\begin{equation}
    c_{ij} =\frac{1}{N_s}\left (   \sum_{j=K+1}^{N_s}\hat{a}_{ij}e^{i \frac{2\pi j}{N_s}} + \sum_{j}^K \hat{b}_{ij}e^{i \frac{2\pi j}{N_s}}\right)
    \label{eq:c_circuit_1}
\end{equation}
Comparing the above two equations leads to the observation that coefficients in $Eq.$ \ref{eq:c_circuit_1} wouldn't be a subset of coefficients in Eq. \ref{eq:c_circuit_2}, and there is no closed-form classical processing/transformation to achieve this. Thus, this parallel circuit results in a somewhat different operation which might be intuitively similar to the sequential circuit, but the output is different. However, Section \ref{sec:Experiments} shows that this conceptually similar operation is also effective in dealing with PDEs/Images and is expected to be more efficient than the sequential circuit for a noisy scenario. 
Also, if we remove the $IQFT$ operation from this circuit and instead apply the classical $IFT$, measuring after Eq.( \ref{eq:3_before_iqft}), we get the following $K$ $N_c\times N_s$ matrices after applying the square root operation:
\begin{equation}
    \left[(\hat{b}^{k})_{k} ,  (\hat{a}^{j})_{j\neq k}\right]_{k=1}^K
\end{equation}
where $\hat{b}^j$ and $\hat{a}^j$ have been defined previously.
In case we combine $\hat{b}^{k}$ from all of the $K$ matrices  with $(\hat{a}^j)_{j\in[K+1, N_s]}$ from any of the $K$ matrices, suppose the first one, it leads to the following $N_c\times N_s$ matrix:
\begin{equation}
    \left[(\hat{b}^{j})_{j\in[1,K]} ,  (\hat{a}^{j})_{j\in[K+1,N_s]}\right]
\end{equation}
which is exactly the same as Eq.(\ref{eq:classical_before_iqft}). Thus, this modified circuit (without the $IQFT$), followed by some classical post-processing and $IFT$, can replicate the classical Fourier layer operation. \\
\textbf{Depth Complexity ($d$).} Given that the only difference compared to the Sequential QFL is the parallel implementation of the controlled parameterised  circuits as against sequential, the depth complexity of this circuit can be derived by substituting $K=1$ in Eq.(\ref{eq:depth_seq}):
\begin{equation}
    \begin{aligned}
        &=  d_{load} + d_{qft} + d_{cpc} + d_{iqft} \\
        &= (N_c+2)\text{log}(N_s) + 3\text{log}(N_c) + N_c \\
    \end{aligned}
    \label{eq:depth_seq}
\end{equation}
and a total of $K$ independent quantum circuits are required to execute this circuit.


\subsection{Composite Circuit for Fourier Layer}
\label{sec:compound_qfno}

\begin{figure}[!htb]
    \centering
    \includegraphics[width=\linewidth]{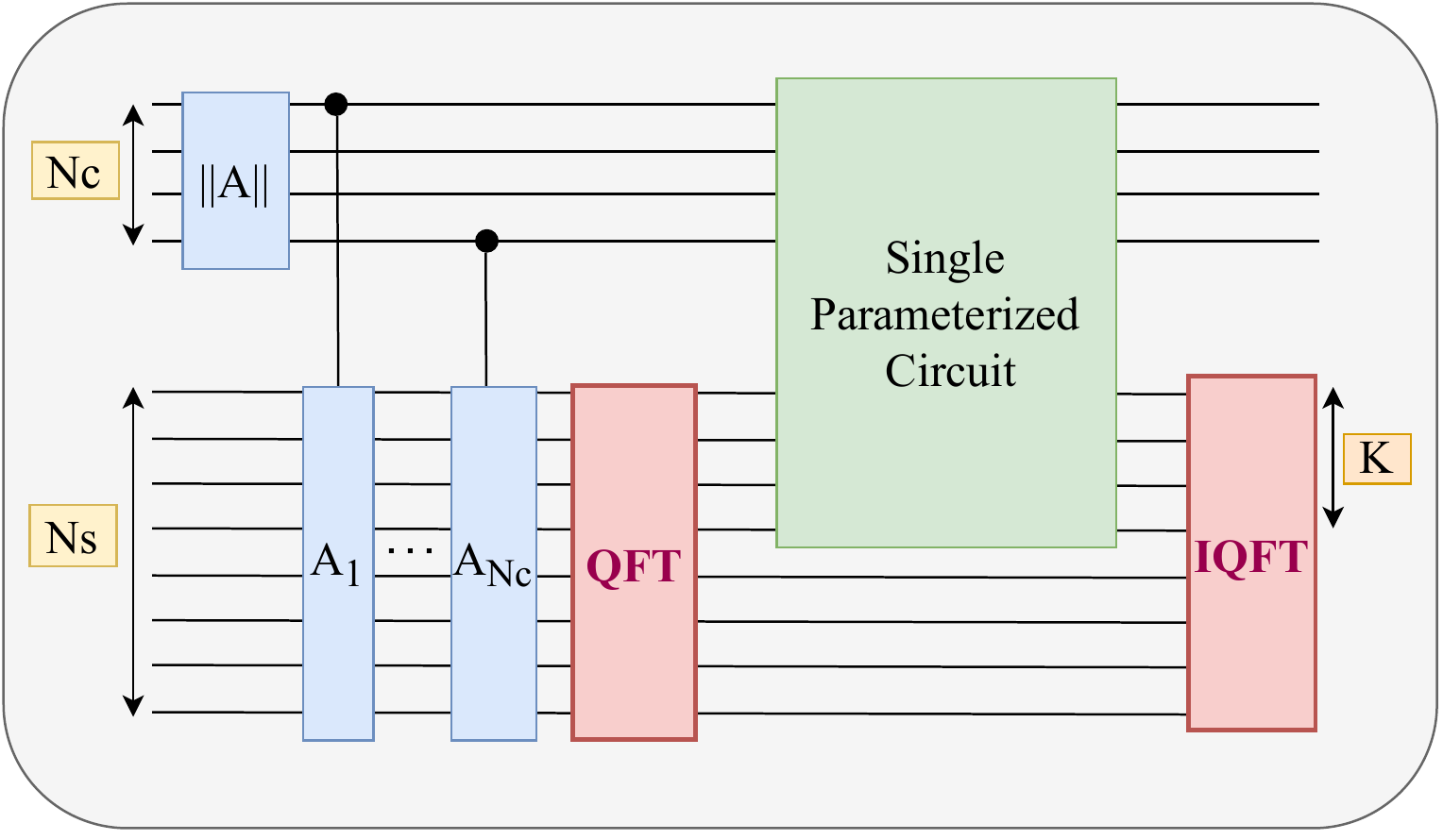}
    \caption{\textbf{Composite-QFL:} Variant of the Sequential-QFNO where instead of controlled butterfly circuits, there is a Composite Butterfly Circuit spanning the upper register and top $K$ qubits of the lower register.}
    \label{fig:compound_qfno_circuit}
\end{figure}


As highlighted in the previous subsection, the depth of the parameterised part of the sequential circuit might make the learning process difficult on currently available noisy quantum hardware. Even though the Parallelised QFL can deal with this, its requirement of $K$ independent $N_c+N_s$ qubit circuits might not be possible in many cases.
Therefore, we propose a new operation corresponding to the learnable part of the sequential circuit and term the resulting overall circuit as the Composite QFL. It significantly decreases the learnable part's depth complexity while requiring only one quantum circuit with ($N_c+N_s$) qubits.
Instead of applying the K-controlled parameterised circuits, we span a single, more extensive parameterised circuit over the upper register ($N_s$ qubits) and top $K$ qubits in the lower register. Fig.\ref{fig:compound_qfno_circuit} shows the diagram for this circuit. 
Note that the upper and lower registers are unary independently, before the parametrised circuit. 

If we jointly consider the upper and lower registers, the states are in a superposition over the hamming weight two basis states. However, we will consider the upper register and just the top $K$ qubits from the lower register, in that case, the states can be in a superposition over hamming weight one and two basis states.
Given that the RBS gates are hamming weight preserving, the state after applying the parameterised circuit will also be a superposition of hamming weight one and two basis states.

Note that the input superposition can't have all the possible basis states with hamming weights 1 and 2 for the top $N_c+K$ qubits. For instance, it doesn't comprise unary states for which the 1 is in the lower $K$ qubits. Similarly, for hamming weight 2, it doesn't comprise states with both the 1s in top $N_c$ or bottom $K$ qubits. In contrast, the output superposition can have any of the hamming weight 1 or 2 states. Recall that the number of possible hamming weight 1 states for these $N_c+K$ qubits is $N_c+K$ and hamming weight 2 states is $N_c+K \choose 2$.

Let us now discuss the application of parameterised circuits on these $N_c+K$ qubits. The complete unitary $B$ will be a $2^{N_c+K} \times 2^{N_c+K}$ block diagonal matrix with each block corresponding to a subspace with fixed hamming weight \cite{kerenidis2022quantum}, $B = B_1 \otimes B_2 \otimes...\otimes B_n$, where $B_i$ correspond to the block diagonal unitary for subspace with hamming weight $i$. Since our input has hamming weight $1$ or $2$, we only care about unitaries $B_1$ and $B_2$. $B_1$ will be of size $(N_c+K) \times (N_c+K)$ and $B_2$ of ${N_c+K \choose 2} \times {N_c+K \choose 2}$.

Given the circuit is similar to the sequential circuit till the $QFT$ operation, the state of this circuit after $QFT$ would be the same as the one in Eq.(\ref{eq:1_after_qft}). 
We now separate this complete state into two sets of states corresponding to hamming weight $1$ and $2$:
\begin{equation}
\begin{aligned}
     \sum_{i}^{N_c}\sum_{j}^{K} \hat{a}_{ij}\ket{e_{i}}\ket{e_{j}} + \sum_{i}^{N_c}\sum_{j=K+1}^{N_s} \hat{a}_{ij}\ket{e_{i}}\ket{e_{j}} \\
\end{aligned}
\label{eq:complete_state_compound_init}
\end{equation}
where the first term corresponds to the superposition of hamming weight two basis states $\ket{h_2}$ and similarly the second term corresponds to the superposition of hamming weight one basis states $\ket{h_1}$.
On application of the parameterised circuit, the unitary $B_1$ will act on $\ket{h_1}$ and $B_2$ on $\ket{h_2}$.\\
Let us first focus on the term corresponding to $\ket{h_1}$. It does not contain the states where the qubits in the upper register are all $0$ and the $1$ lies in the top $K$ qubits of the lower register. It implies that the coefficients of all these states should be taken as zero. Therefore, the state corresponding to this $\ket{h_1}$ can also be written as:
\begin{equation}
    \ket{h_1} = \sum_{i}^{N_c}\sum_{j=K+1}^{N_s} \hat{a}_{ij}\ket{e_{i}}\ket{e_{j}}+\sum_{i}^{K}\sum_{j=K+1}^{N_s} 0\ket{e_{0}}\ket{e_{ij}}
    \label{eq:h1_final}
\end{equation}
where $\ket{e_0}$ denotes the state corresponding to no ones in the upper $N_c$ register and $\ket{e_{ij}}$ denotes the hamming weight 2 states for the lower register, where $i$ and $j$ denote the positions of $1$. Similarly, if we consider the first term in Eq.(\ref{eq:complete_state_compound_init}), corresponding to $\ket{h_2}$, we further have to include states where both ones are in the upper register or both ones in the top $K$ of the lower register. These new states again would have zero coefficients. As a result, we can write the term corresponding to $\ket{h_2}$ in Eq.(\ref{eq:complete_state_compound_init}) as:
\begin{equation}
\begin{aligned}
    \ket{h_2} &= \sum_{i}^{N_c}\sum_{j>i}^{N_c} 0\ket{e_{ij}}\ket{e_{0}}\\
    &+ \sum_{i}^{N_c}\sum_{j}^{K}  \hat{a}_{ij}\ket{e_{i}}\ket{e_{j}}
    +\sum_{i}^{{K}}\sum_{j>i}^{K} 0\ket{e_{0}}\ket{e_{ij}}
    \label{eq:h2_final}
    \end{aligned}
\end{equation}
This results in a total of ${N_c+K \choose 2}$ states.

\noindent
Let us now discuss the application of parameterised circuit ($B_1$, $B_2$) on the $\ket{h_1}$ and $\ket{h_2}$ states in Eq.(\ref{eq:h1_final}) and Eq.(\ref{eq:h2_final}) respectively. For the hamming weight 1 basis, the application of parameterised circuit ($B_1$) is already discussed in sec. \ref{sec:sequential_qfno}. For notational consistency, we denote this operation as a multiplication with matrix $W^{1}\in\mathbb{R}^{(N_c+K)\times(N_c+K)}$. It results in the transformed coefficients $\hat{b}_{ij}$:
\begin{equation}
\begin{aligned}
    \hat{b}_{ij} &= \sum_{t}^{N_c}(W^1_{it}\hat{a}_{tj}) + \sum_{t=N_c+1}^{N_c+K}(W^1_{it}\times0) \; \\&  i \in [1,N_c+K] \;  j \in [K+1,N_s]
    \end{aligned}
\end{equation}
Furthermore, we also apply a post-select operation to preserve the basis, selecting only the states with a non-zero coefficient before applying the $B_1$. 

On similar lines as hamming weight 1 basis, for the hamming weight 2 case, the application of parameterised circuit ($B_2$) can be interpreted as multiplication by the matrix $W^2\in\mathbb{R}^{q\times q}$ where $q={N_c+K\choose2}$. 
Based on a recent work on subspace states \cite{kerenidis2022quantum}, if the parameterised circuit has a nearest neighbour connectivity, then the matrix $B_2$ is the \textit{compound order 2 matrix} \cite{horn2012matrix} of $B_1$. Therefore, for this case, each of its elements will correspond to the determinant of a $2\times2$ submatrix of $W^1$:
\begin{equation}
    W^2_{i,j} = W^1_{a,b}W^1_{a+k,b+k} - W^1_{a+k,b}W^1_{a,b+k}
\end{equation}
for some $a,b,k<N_c+K$. For the butterfly-shaped circuit, we are not limited to nearest neighbour connectivity and thus, $W^2$ has to be extracted from the complete unitary ($2^{N_c+K} \times 2^{N_c+K}$) only. 
After applying this unitary $B_2$ on $\ket{h_2}$, their transformed coefficients $c_{ij}$ are: 
\begin{equation}
\begin{aligned}
    c_{ij} &= \sum_{t}^{N_c\choose2}\left(W^2_{it}\times0\right) + \sum_{t}^{N_cK}\left(W^2_{i,t+{N_c\choose2}}\times a^f_{tj}\right)\\
    &+ \sum_{t=N_cK+{N_c\choose2}}^{N_c+K\choose2}\left(W^2_{it}\times0\right)
    \label{eq:coeff_after_h2}
    \end{aligned}
\end{equation}
Similar to the case of hamming weight 1, here also we use a post-select operation to discard the states which initially had coefficients zero thereby preserving the basis. 

Combining the transformed $\ket{h_1}$, $\ket{h_2}$ states and applying $IQFT$ on the lower register, the final output state of this circuit is:
\begin{equation}
    \sum_{i}^{N_c}\ket{e_i}IQFT\left(\sum_{j}^{K}c_{ij}\ket{e_j} + \sum_{j=K+1}^{N_s}b_{ij}\ket{e_j}\right)
\end{equation}

We term the overall circuit, when parameterised circuit has nearest neighbour connectivity (pyramid-shaped), as the \textit{Composite Circuit (Compound)} having depth complexity $(N_c+K)+\text{log}(N_c)+(N_c+2)\text{log}(N_s)$ and for butterfly shaped as the \textit{Composite Circuit (Butterfly)} having depth complexity $\text{log}(N_c+K)+\text{log}(N_c)+(N_c+2)\text{log}(N_s)$.

\begin{figure*}[!htb]
      \begin{subfigure}[t]{0.47\textwidth}
    \begin{tikzpicture}
    \begin{axis}[
        /pgf/number format/.cd,
        1000 sep={},
        title = {Burgers Equation},
        xlabel={Resolution},
        ylabel={Relative Error},
        legend cell align={left},
        xmin=0, xmax=9000,
        ymin=-0, ymax= 0.42,
        xtick = {2048,4096,6144,8192},
        ytick={ 0.1, 0.2, 0.3, 0.4},
        legend style={at={(1.0,0.6)},anchor=east},
        ymajorgrids=true,
        grid   = major,
        height = 5.2cm,
        width  = 7.65cm
    ]
    \addplot[
        color=blue,
        mark=square*,
        ]
        coordinates {
     (256,   0.04)
     (1024,   0.044)
     (2048,   0.05)
     (4096,   0.035)
     (8192,   0.037)
    };
 
    \addplot[
        color=magenta,
        mark=triangle*,
        ]
        coordinates {
     (256,   0.042)
     (1024,   0.036)
     (2048,   0.041)
     (4096,   0.044)
     (8192,   0.049)
    };
    \addplot[
        color=green,
        mark=triangle*,
        ]
        coordinates {
     (256,   0.074)
     (1024,   0.072)
     (2048,   0.060)
     (4096,   0.057)
     (8192,   0.068)
    };
    \addplot[
        color=red,
        mark=diamond*,
        ]
        coordinates {
     (256,   0.059)
     (1024,   0.069)
     (2048,   0.052)
     (4096,   0.063)
     (8192,   0.057)
    };
    \addplot[
        color=orange,
        mark=diamond*,
        ]
        coordinates {
     (256,   0.17)
     (1024,   0.22)
     (2048,   0.36)
     (4096,   0.39)
     (8192,   0.41)
    };
    \legend{Classical FNO, Sequential \textbf{Q}FNO, Parallel \textbf{Q}FNO , Composite \textbf{Q}FNO, Classical CNNs}
    \end{axis}
    \end{tikzpicture}
    \label{fig:burger}
    \end{subfigure}
           \begin{subfigure}[t]{0.47\textwidth}
    \begin{tikzpicture}
    \begin{axis}[
        title = {Darcy's Flow Equation},
        xlabel={Resolution},
        ylabel near ticks,
        legend cell align={left},
        xmin=0, xmax=480,
        ymin=-0, ymax= 1.5,
        xtick= {80, 160, 240, 320, 400, 480},
        ytick={0.3, 0.6, 0.9, 1.2, 1.5},
        legend style={at={(1.0,0.6)},anchor=east,fill=none},
        ymajorgrids=true,
        grid   = major,
        height = 5.2cm,
        width  = 7.65cm
    ]
    \addplot[
        color=blue,
        mark=square*,
        ]
        coordinates {
     (85,   0.13)
     (141,   0.17)
     (211,   0.21)
     (331, 0.19)
     (421,   0.16)
    };
 
    \addplot[
        color=magenta,
        mark=triangle*,
        ]
        coordinates {
     (85,   0.10)
     (141,   0.12)
     (211,   0.24)
     (331, 0.23)
     (421,   0.18)
    };
    \addplot[
        color=green,
        mark=triangle*,
        ]
        coordinates {
     (85,   0.21)
     (141,   0.31)
     (211,   0.35)
     (331, 0.32)
     (421,   0.31)
    };
    \addplot[
        color=red,
        mark=diamond*,
        ]
        coordinates {
     (85,   0.18)
     (141,   0.24)
     (211,   0.29)
     (331, 0.27)
     (421,   0.24)
    };
    \addplot[
        color=orange,
        mark=diamond*,
        ]
        coordinates {
     (85,   0.32)
     (141,   0.64)
     (211,   0.89)
     (331, 0.94)
     (421,   1.2)
    };
    \end{axis}
    \end{tikzpicture}
    \label{fig:burger}
    \end{subfigure} 
\caption{ \textbf{Left}: Performance comparison (relative error as used in \cite{li2020fourier}) of the classical Fourier networks, CNNs and the three circuit proposals for a quantum Fourier layer on the Burgers 1D PDE equation across different resolutions. The quantum Fourier circuits are quite close to the performance of the classical Fourier baseline and much better than classical CNNs in minimising the error. \textbf{Right}: Same comparison on the Darcy's 2D PDE for different resolutions. A similar relative performance is observed where the error in CNNs is much larger and is increasing w.r.t. resolution, whereas the error in the other four (3 quantum circuits and classical layer) is quite similar.}
\label{fig:pde_plots}
\end{figure*}
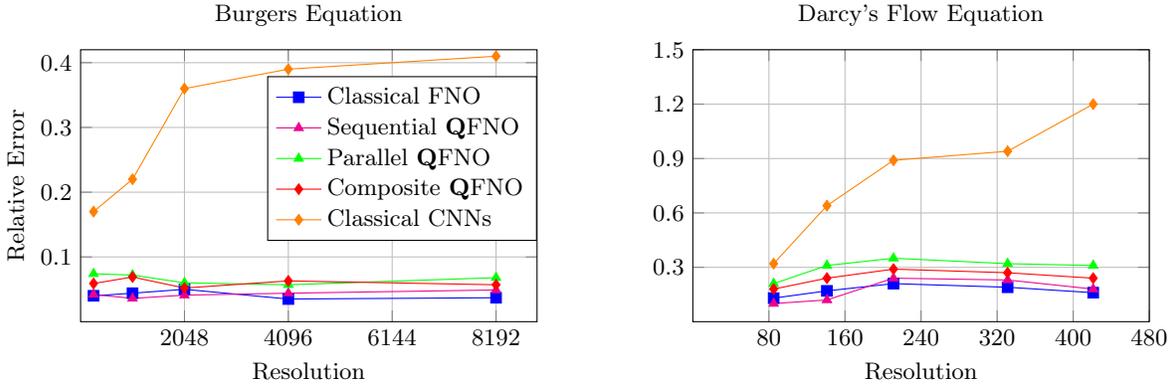

\section{Experiments}
\label{sec:Experiments}
This section analyses our proposed quantum algorithms for solving PDEs and image classification tasks. We compare them against the \textit{state-of-the-art} in both the domains, \textit{i.e.}, classical Fourier Networks (for PDEs) and CNNs (for image classification), for both tasks. All the details related to architecture and hyperparameters are provided at the end of the paper. All the experiments shown in this section are simulated, \textit{i.e.}, the quantum operations have been simulated using classical matrices corresponding to quantum unitaries, since the currently available quantum hardware is too noisy for circuits of such size. However, we expect the upcoming generations of quantum hardware to support experiments of such scale.
Also, the Butterfly version of the Composite Circuit, \textit{i.e.} \textit{Composite Circuit (Butterfly)}, has been used for all the experiments.

\subsection{Partial Differential Equations}

We show results on all the three PDEs used in the classical Fourier Layer paper \cite{li2020fourier}: Burgers' equation, Darcy's Flow equation and Navier-Stokes equation using the datasets proposed in that paper. All of these equations were designed for modelling the flow of fluids and have found their applications in other domains as well. We abstractly describe the three tasks. For equations and other details, please refer to \cite{li2020fourier}. We analyse the performance of the trained networks across different resolutions ($N_s$) for the first two and for different viscosity values for the third.  \\
\textbf{Burgers' Equation}. It is a 1D-Partial Differential Equation for modelling fluid motion and is expressed as follows:
\begin{equation}
\begin{aligned}
    \partial_t u(x,t) + \partial_x (u^2(x,t)/2) &= \nu\partial_{xx}u(x,t) & t\in(0,1] \\
    \text{where} \quad u(x,0) &= u_0(x) \quad\text{and}  & x\in(0,1)
\end{aligned}
\end{equation}
with $\nu\in\mathbb{R}_+$ corresponding to the fluid viscosity and $u_0$ denoting the initial condition function for this PDE family. 
We need to learn the mapping from this $u_0$ to the function at time one $u(x,1)$, for a given viscosity.

\begin{figure}[!htb]
    \centering
    \includegraphics[width=0.98\linewidth]{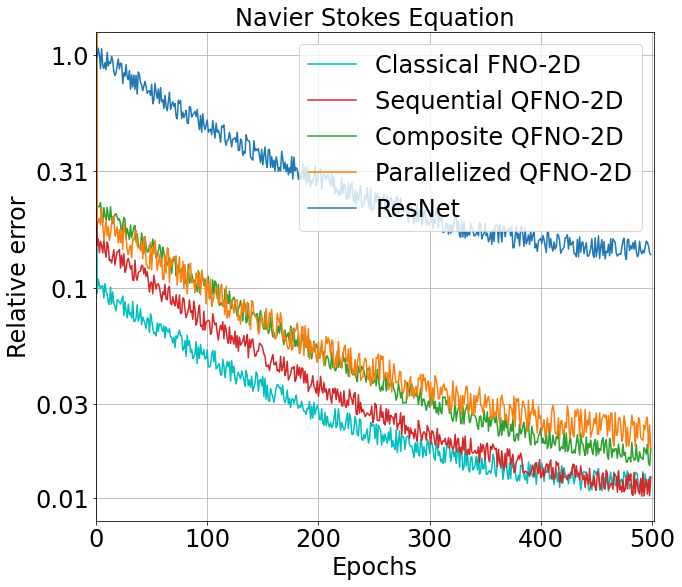}
    \caption{Convergence comparison for the Navier-Stokes equation with $v=1e-3$, trained for $500$ epochs. }
    \label{fig:navier}
\end{figure}


\begin{figure*}[!htb]
      \begin{subfigure}[t]{0.36\textwidth}
    \begin{tikzpicture}
    \begin{axis}[
        title = {MNIST},
        xlabel={Epochs},
        ylabel={Accuracy},
        legend cell align={left},
        xmin=0, xmax=10,
        ymin=50, ymax= 100,
        xtick={ 2, 4, 6, 8, 10},
        ytick={60, 70, 80, 90, 100},
        legend style={at={(1.0,0.22)},anchor=east, nodes={scale=0.8}},
        ymajorgrids=true,
        grid   = major,
        height = 6.2cm,
        width  = 6.25cm
    ]
    \addplot[
        color=blue,
        mark=square*,
        ]
        coordinates {
     (1,85.7)
     (2,87.98)
     (3,89.7)
     (4,93.5)
     (5,95.7)
     (6,96.2)
     (7,96.4)
     (8,96.5)
     (9,97.1)
     (10,97.1)
    };
 
    \addplot[
        color=magenta,
        mark=triangle*,
        ]
        coordinates {
     (1,88.7)
     (2,90.98)
     (3,92.7)
     (4,93.5)
     (5,94.7)
     (6,96.2)
     (7,97.1)
     (8,98.1)
     (9,98.5)
     (10,98.7)
    };    
    \addplot[
        color=green,
        mark=triangle*,
        ]
        coordinates {
      (1,87.2)
     (2,88.98)
     (3,90.7)
     (4,91.1)
     (5,94.3)
     (6,95.2)
     (7,97.1)
     (8,97.4)
     (9,97.6)
     (10,97.9)
    };
    \addplot[
        color=red,
        mark=diamond*,
        ]
        coordinates {
     (1,86.7)
     (2,89.98)
     (3,91.7)
     (4,92.1)
     (5,93.9)
     (6,96.2)
     (7,96.9)
     (8,97.5)
     (9,97.9)
     (10,98.2)
    };
    [90.21, 91.87, 92.36, 95.6, 96.8, 97.8, 98.2, 98.6, 99, 99.2]
    \addplot[
        color=orange,
        mark=diamond*,
        ]
        coordinates {
     (1,90.21)
     (2,91.87)
     (3,92.36)
     (4,95.6)
     (5,96.8)
     (6,97.8)
     (7,98.2)
     (8,98.6)
     (9,99)
     (10,99.2)
    };
    \legend{Classical FNO, Sequential \textbf{Q}FNO, Parallel \textbf{Q}FNO , Composite \textbf{Q}FNO, Classical CNNs}
    \end{axis}
    \end{tikzpicture}
    \label{fig:burger}
    \end{subfigure}
\begin{subfigure}[t]{0.29\textwidth}
    \begin{tikzpicture}
    \begin{axis}[
        title = {Fashion-MNIST},
        xlabel={Epochs},
        ylabel near ticks,
        legend cell align={left},
        xmin=0, xmax=10,
        ymin=75, ymax= 100,
        xtick={ 2, 4, 6, 8, 10},
        ytick={80, 85, 90, 95, 100},
        ymajorticks=false,
        legend style={at={(1.0,0.22)},anchor=east,fill=none, nodes={scale=0.8}},
        ymajorgrids=true,
        grid   = major,
        height = 6.2cm,
        width  = 6.45cm
    ]
    \addplot[
        color=blue,
        mark=square*,
        ]
        coordinates {
     (1,80.14)
     (2,81.98)
     (3,81.78)
     (4,82.5)
     (5,83.7)
     (6,85.2)
     (7,88.4)
     (8,89.5)
     (9,89.9)
     (10,90.1)
    };
    \addplot[
        color=magenta,
        mark=triangle*,
        ]
        coordinates {
     (1,82.2)
     (2,83.98)
     (3,84.78)
     (4,85.1)
     (5,86.2)
     (6,88.2)
     (7,89.1)
     (8,91.2)
     (9,91.5)
     (10,91.7)
    };
    \addplot[
        color=green,
        mark=triangle*,
        ]
        coordinates {
     (1,80.7)
     (2,81.98)
     (3,84.1)
     (4,83.9)
     (5,84.9)
     (6,87.5)
     (7,89.9)
     (8,90.5)
     (9,90.9)
     (10,91.1)
    };
    \addplot[
        color=red,
        mark=diamond*,
        ]
        coordinates {
     (1,79.7)
     (2,80.98)
     (3,81.1)
     (4,82.4)
     (5,83.9)
     (6,85.5)
     (7,88.3)
     (8,89.9)
     (9,90.7)
     (10,90.9)
    };
    \addplot[
        color=orange,
        mark=diamond*,
        ]
        coordinates {
     (1,84.2)
     (2,85.87)
     (3,86.36)
     (4,86.6)
     (5,87.8)
     (6,89.1)
     (7,91.4)
     (8,91.9)
     (9,92.0)
     (10,92.1)
    };
    \end{axis}
    \end{tikzpicture}
    \label{fig:burger}
    \end{subfigure} 
           \begin{subfigure}[t]{0.33\textwidth}
    \begin{tikzpicture}
    \begin{axis}[
        title = {Pneumonia-MNIST},
        xlabel={Epochs},
        ylabel near ticks,
        legend cell align={left},
       xmin=0, xmax=15,
        ymin=50, ymax= 100,
        xtick={ 3, 6, 9, 12, 15},
        ytick={60, 70, 80, 90, 100},
        ymajorticks=false,
        legend style={at={(1.0,0.6)},anchor=east,fill=none},
        ymajorgrids=true,
        grid   = major,
        height = 6.2cm,
        width  = 6.55cm
    ]
    \addplot[
        color=blue,
        mark=square*,
        ]
        coordinates {
     (1,52.1)
     (2,61.98)
     (3,68.78)
     (4,76.5)
     (5,83.7)
     (6,80.2)
     (7,82.4)
     (8,83.5)
     (9,81.2)
     (10,82.4)
     (11,83.7)
     (12,84.1)
     (13,83.6)
     (14,84.3)
     (15,84.4)
    };
    \addplot[
        color=magenta,
        mark=triangle*,
        ]
        coordinates {
     (1,59.2)
     (2,65.98)
     (3,72.78)
     (4,77.1)
     (5,79.9)
     (6,82.2)
     (7,85.1)
     (8,86.3)
     (9,86.1)
     (10,83.3)
     (11,85.7)
     (12,86.3)
     (13,86.7)
     (14,86.3)
     (15,86.9)
    };
    \addplot[
        color=green,
        mark=triangle*,
        ]
        coordinates {
     (1,56.2)
     (2,61.98)
     (3,66.78)
     (4,72.1)
     (5,76.9)
     (6,81.2)
     (7,81.5)
     (8,81.6)
     (9,80.1)
     (10,81.9)
     (11,82.7)
     (12,84.1)
     (13,84.9)
     (14,85.4)
     (15,85.8)
    };
    \addplot[
        color=red,
        mark=diamond*,
        ]
        coordinates {
     (1,53.2)
     (2,62.98)
     (3,68.78)
     (4,74.1)
     (5,72.9)
     (6,79.2)
     (7,80.1)
     (8,81.3)
     (9,78.1)
     (10,82.3)
     (11,84.7)
     (12,85.1)
     (13,84.6)
     (14,85.0)
     (15,85.1)
    };
    \addplot[
        color=orange,
        mark=diamond*,
        ]
        coordinates {
     (1,67.48)
     (2,75.98)
     (3,81.38)
     (4,73.37)
     (5,78.8)
     (6,82.47)
     (7,83.58)
     (8,79.76)
     (9,85.67)
     (10,87.88)
     (11,84.38)
     (12,85.78)
     (13,86.9)
     (14,87.8)
     (15,88.1)
    };
    \end{axis}
    \end{tikzpicture}
    \label{fig:burger}
    \end{subfigure} 
\vspace{-7.0mm}
\caption{\textbf{Left}: Performance comparison of the CNNS, classical Fourier layer and the proposed quantum circuits on the MNIST dataset. It can be observed that all of them perform quite similarly, classical CNNs being the best. \textbf{Middle}: A similar comparison on the Pneumonia-MNIST \cite{yang2021medmnist} dataset. The performance of CNNs is somewhat noisy here whereas it is smoother in the case of the sequential circuit, both converging to a similar value. The composite quantum circuit and the classical Fourier baseline are also quite close to the CNNs in convergence. \textbf{Right}: Same comparison on the FashionMNIST \cite{xiao2017fashion} data. Here, a significant difference in the performance is observed with CNNs being the best followed by the Sequential circuit.}
\label{fig:classification_plots}
\end{figure*}
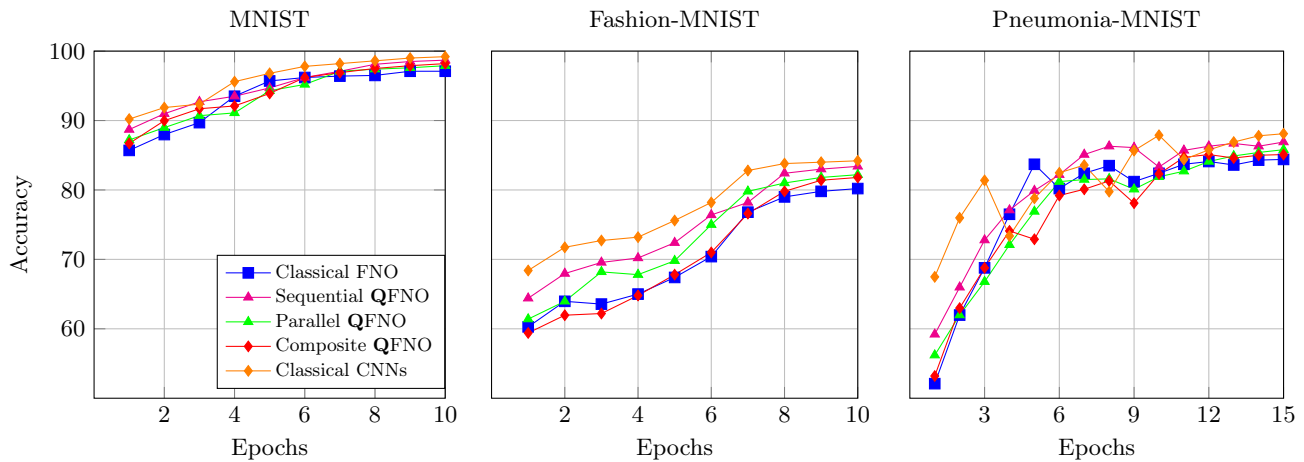

\begin{table*}[!htb]
    \centering
    \vskip 0.15in
    \begin{center}
    \begin{small}
    \begin{sc}
    \resizebox{\linewidth}{!}{
    \begin{tabular}{lccccr}
    \toprule
        Method  & Classical FNO & Sequential QFNO & Parallelised QFNO & Composite  QFNO \\
        Parameters & 294,912 & 23,040 & 23,040 & 6,144 \\
        \midrule
        $\nu=1e-3$; $T=50$ & 0.0139 & 0.0148 & 0.0167 & 0.0186 \\
        $\nu=1e-4$; $T=30$ & 0.1603 & 0.1618 & 0.1633 & 0.1660 \\
        $\nu=1e-5$; $T=20$ & 0.1601 & 0.1615 & 0.1626 & 0.1638 \\
    \bottomrule
    \end{tabular}
    }
    \end{sc}
    \end{small}
    \end{center}
    \vspace{-0.2in}
    \caption{ Comparison of parameters required by one layer of the proposed circuits and the existing classical Fourier Layer along with error analysis for different $\nu$ and $T$ values for the 2D case of a Navier-Stokes equation. 
    }
    \label{tab:navier}
\end{table*}

Fig.\ref{fig:pde_plots} (left) shows the comparison of relative error in estimating this mapping among the classical Fourier Layer, classical CNNs and proposed quantum circuits for the Fourier Layers, across different resolutions. The quantum circuits perform comparably to the classical Fourier Layer and are much better than classical CNNs. \\ 
\textbf{Darcy's Flow Equation}.
In this case, it is a 2D PDE with the following equation:
\begin{equation}
\begin{aligned}
    -\nabla\cdot\left(a(x)\nabla u(x)\right) &= f(x) \qquad x\in(0,1)^2\\
    u(x) &= 0 \qquad \quad x\in \partial(0,1)^2
\end{aligned}    
\end{equation}
where $a(x)$ is the diffusion coefficient, $f(x)$ is the forcing function and $u(x)$ is the solution function. The aim here is to learn the mapping $a\mapsto u$ given the forcing function $f(x)$. All of them are functions of positional coordinates only. 
Fig.\ref{fig:pde_plots} (right) shows the relative error for the 2D-version of all the methods in solving this PDE, across different resolutions. Here also the three quantum circuits and the classical Fourier layer show similar performance, consistent across resolutions, and the CNNs show much worse results, with their error increasing with resolution. \\
\textbf{Navier-Stokes Equation}. We now consider the 2D Navier-Stokes Equation which is as follows:
\begin{equation}
\begin{aligned}
    \partial_t w(x&,t) + u(x,t)\cdot\nabla w(x,t) = \nu\delta w(x,t) + f(x), \\
    &\nabla\cdot u(x,t)=0 \qquad w(x,0) = w_0(x)\\
    &\quad \quad x \in (0,1)^2 \qquad t\in(0,T]
\end{aligned}
\end{equation}
where $w$ corresponds to vorticity, $w_0$ being the initial vorticity,  $\nu$ is the viscosity, $u$ is  a velocity field and $f(x)$ is some sort of a forcing function. The aim here is to model the fluid vorticity up to instant $T(>10)$ given the vorticity up to time $10$.
Fig.\ref{fig:navier} shows the performance comparison for this equation between our proposed circuits and classical methods. It shows the convergence comparison for this family with viscosity $\nu$ fixed to $1e-3$ for all the methods. Here, again, it can be observed that all the proposed circuits and the classical Fourier method perform significantly better than CNNs. Also, from Table \ref{tab:navier}, the sequential circuit performs similarly to classical method  and the others converge at a slightly higher error.

\subsection{Image Classification}

We further compare our proposed Quantum algorithms on the downstream image classification tasks on benchmark datasets including the MNIST, FashionMNIST \cite{xiao2017fashion} and PneumoniaMNIST \cite{yang2021medmnist} datasets. The MNIST dataset consists of 
grayscale images corresponding to digits from 0 to 9 (both included),
having a resolution of $28\times 28$. The task is to predict the digit for a given input image. Similarly, for FashionMNIST also there is a 10-way classification task into various categories of clothing, using $28\times 28$ grayscale images. On the other hand, PneumoniaMNIST involves a binary classification task into positive or negative for a given grayscale image.

For this comparison, we have used the 2D versions of all the architectures--our quantum FNO, the classical FNO and the CNNs.
Fig.\ref{fig:classification_plots} shows the epochs (x-axis) v/s accuracy (y-axis) plot for this evaluation. It can be observed that our proposed algorithms outperform the classical FNO and converge in close proximity (w.r.t. accuracy) to the classical CNNs, with Sequential QFL being almost comparable to the classical CNNs. This shows that the proposed QFLs are comparable in performance to the \textit{state-of-art} in the vision domain as well along with solving PDEs, thereby broadening the scope of their applicability.

\section{Conclusion}
We proposed a quantum algorithm to carry out the recently proposed classical Fourier Neural Operator on quantum hardware. We further proposed two more quantum algorithms, which perform a different operation than the classical one and can be much more efficiently deployed on noisy quantum hardware. Experimental results further confirm that the proposed quantum neural networks perform efficiently in both solving PDEs and image classification. The sequential network matches the best-performing classical algorithm (the CNNs for images and the classical Fourier Layer for PDEs) on both tasks. 

An interesting future direction can be further developing the learning process of the composite network so that it can perform better than the sequential network while at the same time being more efficient to deploy. The composite network intuitively performs a kind of attention mechanism that learns how to mix the mappings performed on each of the top $K$ modes, and understanding how this mixing works can provide new ideas for improving it.

\bibliographystyle{alpha}
\bibliography{egbib}

\end{document}